\documentclass[a4paper,11pt]{article}
\usepackage{amsmath, amssymb}

\usepackage{jcappub}
\usepackage{newtxtext,newtxmath}
\usepackage{graphicx}
\usepackage{hyperref}
\DeclareRobustCommand{\VAN}[3]{#2}
\let\VANthebibliography\thebibliography
\def\thebibliography{\DeclareRobustCommand{\VAN}[3]{##3}\VANthebibliography}

\newcommand{\Lya}{Lyman-$\alpha$}
\newcommand{\astrid}{\texttt{ASTRID}}

\newcommand{\mpgadget}{{\small MP-GADGET}}

\title{{PRIYA:} A New Suite of Lyman-$\alpha$ Forest Simulations for Cosmology}

\author[a]{Simeon Bird,}
\affiliation[a]{Department of Physics \& Astronomy, University of California  Riverside,\\900 University Avenue, Riverside, CA 92521, USA}
\author[a]{Martin Fernandez,}
\author[a]{Ming-Feng Ho,}
\author[a]{Mahdi Qezlou,}
\author[a]{Reza Monadi,}
\author[b]{Yueying Ni, }
\author[c]{Nianyi Chen,}
\author[c,d]{Rupert Croft, }
\author[c,d]{Tiziana Di Matteo.}
\affiliation[b]{Harvard-Smithsonian Center for Astrophysics, 60 Garden Street, Cambridge, MA 02138, USA}
\affiliation[c]{McWilliams Center for Cosmology, Department of Physics, Carnegie Mellon University, Pittsburgh, PA 15213}
\affiliation[d]{NSF AI Planning Institute for Physics of the Future, Carnegie Mellon University, Pittsburgh, PA 15213, USA}
\emailAdd{sbird@ucr.edu}

\abstract{
We present the PRIYA suite of cosmological simulations, based on the code and hydrodynamic model of the ASTRID simulation, and designed for cosmological analyses of the Lyman-$\alpha$ forest. Our simulation suite spans a $9$-dimensional parameter space, including $4$ cosmological parameters and $5$ astrophysical/thermal parameters. We have run $48$ low fidelity simulations with $1536^3$ particles in a $120$ Mpc/h box and $3$ high fidelity simulations with $3072^3$ particles in a $120$ Mpc/h box. All our simulations include a full physics model for galaxy formation, including supernova and AGN feedback, and thus also contain a realistic population of DLAs. We advance on earlier simulations suites by larger particle loads, by incorporating new physical models for patchy hydrogen and helium reionization, and by self-consistently incorporating a model for AGN feedback. We show that patchy helium reionization imprints an excess in the 1D flux power spectrum on large scales, which may allow future measurements of helium reionization bubble sizes.
Simulation parameters are chosen based on a Latin hypercube design and a Gaussian process is used to interpolate to arbitrary parameter combinations.
We build a multi-fidelity emulator for the 1D flux power spectrum and the mean IGM temperature. We show that our final interpolation error is $< 1\%$ and that our simulations produce a flux power spectrum converged at the percent level for $z=5.4$ - $2.2$. Our simulation suite will be used to interpret Lyman-$\alpha$ forest 1D flux power spectra from SDSS and future DESI data releases.
}

\begin{document}

\maketitle

\section{Introduction}


The distribution of neutral hydrogen in the Universe, as traced by \Lya~forest absorption in quasars \cite{Gunn:1965}, is one of the classic probes of cosmological structure \citep{Croft:1998}. The power of the \Lya~forest lies in its ability to probe gravitational collapse on small scales at $z = 2-5$ in a way which reduces sensitivity to the uncertain astrophysics of galaxy formation \citep{Cen:1994,Zhang:1995,Miralda-Escude:1996,Hernquist:1996,Rauch:1997}. However, realising this potential requires cosmological simulations, which are the only way to model the non-linear evolution of the gas and dark matter with sufficient fidelity.
Creating an accurate simulation of the \Lya~forest requires the ability to probe the pressure smoothing scale of the collapsing gas and thus, when including a temperature boost from reionization, a mean interparticle spacing of $\sim 40$ kpc/h \cite{Borde:2014}.
 Conversely, suppressing cosmic variance in the simulation requires a box size of $\sim 100$ Mpc/h.
These challenging simulation requirements mean that the constraining power of the \Lya~forest is in some respects limited by our ability to create large and accurate enough simulation models.

Here we present the PRIYA suite of simulations, which are run using the \mpgadget~code developed for the ASTRID project \cite{Ni:2021, Bird:2022}. Our largest simulations have a box size of $120$ Mpc/h and a mean interparticle spacing of $39$ kpc/h, including $2\times 3072^3$ particles, and meeting the criteria for resolving the \Lya~forest.
We show that we can model scales from $k \sim 2\times 10^{-3}$ - $0.1$ s/km and redshifts $2.2 - 5.4$ to an accuracy of a few percent, as required by current survey data. Our good resolution convergence relies to some degree on our reionization model, as discussed in Section~\ref{sec:reionresolution}. In addition, the simulations contain a full hydrodynamic galaxy formation model including stellar and black hole physics. The main output is the \Lya~spectra, of which we have generated three grids, one for each axis, of $480^2$ spectra each, for a total of $691,200$ spectra from each snapshot. We also report the derived 1D flux power spectra, and build an emulator for it as a function of cosmology. We are able to predict the output flux power spectrum from a simulation to an accuracy of a few percent over a wide range of cosmological parameters.

There have been a variety of previous \Lya~forest simulation suites. Ref.~\cite{McDonald:2005pk} used dark matter only simulations convolved with a temperature broadening function. Ref.~\cite{Viel:2006} used hydrodynamic simulations with a simplified star formation prescription. Each cosmological parameter was varied individually and a polynomial fit to the changes in the flux power spectrum. This approach was extended by Ref.~\cite{Borde:2014, Rossi:2020} with larger simulations, necessitated by the more precise dataset available in later SDSS releases. Ref.~\cite{Bird:2019} improved on the polynomial fit with Gaussian Process emulator techniques for the \Lya~forest. However, the simulations used were small and did not fully resolve the forest. Ref.~\cite{Walther:2019} also used an emulator to study the evolution of the IGM temperature. Ref.~\cite{Pedersen:2021, Pedersen:2022} pursued a complementary approach, building an emulated map between the flux power spectrum and the amplitude and slope of the matter power spectrum at $z\sim 3$. Ref~\cite{Esposito:2022} pursued a similar approach to model small scales in the flux power spectrum, and Ref.~\cite{Cabayol:2023} used a neural network emulator. There have also been several large-scale \Lya~simulation projects. The Sherwood simulations \cite{Bolton:2017} created extremely high resolution models of the $z=5$ intergalactic medium. Ref.~\cite{Puchwein:2022} ran a suite of simulations, SHERWOOD-RELICS, to model the interaction between the intergalactic medium and hydrogen reionization, using a novel two-step procedure to include the effect of inhomogeneous reionization with radiative transfer. Ref.~\cite{Hadzhiyska:2023} generated a single mock on Gpc scales by applying the Fluctuating Gunn-Peterson Approximation to a large dark matter only simulation.

The main improvements in the simulation suite we present here come from our larger boxes coupled with a high effective resolution, as well as our inclusion of patchy hydrogen and helium reionization models. We double the effective resolution of our emulator by using the multi-fidelity technique from Ref.~\cite{Ho:2022}, as applied to the \Lya~forest by Ref.~\cite{Fernandez:2022}. Finally, we use a full-physics galaxy formation model incorporating self-shielding, star formation and stellar and AGN feedback. We perform a large suite of simulations which vary both cosmological and astrophysical parameters simultaneously. Overall, we are able to model the flux power spectrum over a wide range of scales and cosmologies at high (percent-level) accuracy.

The largest \Lya~forest dataset is that from Sloan Digital Sky Survey (SDSS) quasars, most recently Data Release (DR) 17, although the latest flux power spectrum estimate uses SDSS DR14 \cite{Chabanier:2019}. SDSS is limited both by low signal to noise and moderate spectral resolution, which does not allow it to probe the thermal cutoff of the intergalactic gas and measure small scales. There are thus independent sets of spectra, observing fewer quasars at higher spectral resolution. These include the XQ-100 survey \cite{Irsic:2017pk, Esposito:2022} at moderate spectral resolution, and the higher resolution KODIAQ/SQUAD \cite{KODIAQ:2022} data, as well as the spectra from HIRES/UVES \cite{Viel:2013wdm, Boera:2019}. These datasets have been used to constrain cosmological parameters and make constraints on total neutrino mass ($\sum m_\nu < 0.11$ eV when combined with the cosmic microwave background) \citep{2004MNRAS.354..684V, McDonald:2005pk, Viel:2006, 2005PhRvD..71j3515S, 2006JCAP...10..014S, Rossi:2017, 2020JCAP...04..038P}, measure the expansion rate (via baryon acoustic oscillations) at $z \sim 2.4$ \cite{Slosar:2011, Bautista:2017, dSAgathe:2019} and test alternatives to cold dark matter \citep{2005PhRvD..71f3534V,  Viel:2013wdm, Irsic:2017pk, Garzilli:2021, 2021PhRvL.126g1302R}. Astrophysical applications include measuring the thermal history of the Universe \citep{2008MNRAS.386.1131B,2014MNRAS.438.2499B, 2016MNRAS.463.2335N, Boera:2019, Gaikwad:2021, Villasenor:2022, Yang:2023}, ionizing background models \cite{Puchwein:2019, FG:2020}, the timing of reionization \citep{Gaikwad:2017,Onorbe:2017,Chardin:2017,DAloisio:2017}, and creating 3D maps of neutral hydrogen density \citep{Lee:2016, Ozbek:2016, Lee:2018, Horowitz:2022, Qezlou:2022}.

In this paper we present a simulation model designed to mock scales probed by SDSS DR14 and the Dark Energy Spectroscopic Instrument (DESI) ($k \lesssim 0.05$ s/km) at the percent level, comparable to the statistical accuracy of these datasets. We will show that we are also able to model smaller scales ($k \lesssim 0.1$ s/km) at the ten-percent level required by the statistical error in XQ-100 \cite{KODIAQ:2022}. 
Section~\ref{sec:simulations} introduces the simulation model, including the generation of spectra and the emulator for the 1D flux power spectrum. We show in Section~\ref{sec:looerror} that the interpolation error in our emulator is small. Section~\ref{sec:boxsize} demonstrates that our simulations produce a converged model for the \Lya~forest. Section~\ref{sec:thermal} shows our derived thermal histories and Section~\ref{sec:singleparams} shows how our simulation parameters affect the 1D flux power spectrum. We conclude in Section~\ref{sec:conclusions}.

\section{Simulations}
\label{sec:simulations}

We have performed cosmological hydrodynamic simulations at two resolutions, over a parameter space containing a total of $9$ parameters describing both cosmology and astrophysics. Our lower resolution simulations, of which we have $48$, contain $1536^3$ gas particles, $1536^3$ cold dark matter particles and model a $120$ Mpc/h periodic box with a mean interparticle spacing of $79$ kpc/h. Our $3$ higher resolution simulations contain $2\times 3072^3$ particles and thus a mean interparticle spacing of $39$ kpc/h. These choices are justified in Section~\ref{sec:boxsize}. All simulations were run to $z=2.2$, the lowest redshift at which SDSS measures the \Lya~forest flux power spectrum.

Simulations were run using \mpgadget~\cite{MPGadget2018, Bird:2022}, a massively scalable version of the cosmological structure formation code Gadget-3 \cite{Springel:2005}. The generation of initial conditions is described in Section~\ref{sec:initconds}. Section~\ref{sec:gravity} describes the gravity solver, hydrodynamic solver, and cooling. Our simulations incorporate physical models for inhomogeneous hydrogen and helium reionization. The parameters of these models are varied in our emulator in order to model uncertainty in the thermal history of the intergalactic medium (IGM).
This differs from several earlier simulation models, which generated arbitrary thermal histories by scaling the heating rates directly \cite[e.g.~][]{Viel:2006, Viel:2013wdm, Irsic:2017, Garzilli:2019}. Scaling the heating rates allows the greatest possible freedom in the redshift evolution of the thermal histories, but does not directly impose a physically plausible thermal history, which can sometimes lead to best-fit thermal histories that are not physically realisable \cite{Walther:2019}.
Using physical reionization models ensures that our simulations all contain physically plausible, self-consistent thermal histories. This choice is enabled by two advances: first, our simulation boxes are large enough to contain many helium reionization bubbles and, second, the thermal history of the IGM is now reasonably well constrained, limiting the variety of models it is necessary to marginalise over.
Sections~\ref{sec:hydrogen} and ~\ref{sec:helium} describe our models for hydrogen and helium reionization, respectively, and how their parameters map to the familiar thermal parameters of the \Lya~forest.

Our simulations include a full galaxy formation model with star formation, stellar winds and AGN feedback. We use these simulations primarily because they self-consistently incorporate our AGN feedback model, avoiding the need for a post-processing correction. AGN feedback has been shown to influence the \Lya~forest flux power spectrum \cite{Viel:2013, Chabanier:2020}. Full-physics simulations also allow us to self-consistently include the effect of self-shielded gas in the modelling, by producing Damped \Lya~Absorbers (DLAs) in our artificial spectra and then masking them in the same way as the observational analysis.
Finally, with our simulation code the computational cost between a full physics simulation and an equivalent simulation with a simplified star formation model (``quick \Lya'') is only $\sim 30\%$, and so the cost is relatively small. Using full physics simulations allows our suite to be used for other applications in future work.

Our galaxy formation model parameters generally follow those used in the \astrid~simulation, as detailed in Refs.~\cite{Bird:2022, Ni:2021}. Here we summarise the main features of the model, and refer the interested reader to \cite{Bird:2022} for more details. Readers familiar with \astrid~should refer to Table~\ref{tab:paramchanges}, which summarises the differences in the galaxy formation model. We describe the star formation and stellar wind model in Section~\ref{sec:stellar}, and the black hole and AGN feedback model in Section~\ref{sec:agn}.

\begin{table*}
\begin{centering}
  \begin{tabular}{lll}
  \hline
  Parameter & PRIYA Value & \astrid~Value \\
    \hline
SPH density kernel  & Cubic & Quintic \\
Minimum wind velocity $v_w$ & $100$ km/s & $0$ \\
Temperature at HI reionization & $15000$ K & $0$ K \\
Stars per gas particle & $1$ & $4$ \\
Metal return from stars & Disabled & Enabled\\
Metal line cooling & Disabled & Enabled. \\
    \hline
  \end{tabular}
  \caption{Summary of changes and simplifications in the galaxy formation model from the \astrid~simulation.}
  \label{tab:paramchanges}
  \end{centering}
\end{table*}


\subsection{Initial Conditions}
\label{sec:initconds}

We generate initial condition for our simulations as in Ref.~\cite{Bird:2020}. We use separate transfer functions for gas and cold dark matter particles, as generated by CLASS \cite{CLASS}. Velocities are initialised using the respective velocity transfer functions, so that the variation in halo gas fraction from dark matter-baryon relative velocities are automatically included in the simulation. Radiation is included in the cosmological background evolution. Simulations are initialised at $z=99$. Cold dark matter is initialised on a regular grid and gas particles are initialised using a force-free Lagrangian glass, using the procedure outlined in Ref.~\cite{Bird:2020}.

The purpose of the simulation suite is to constrain cosmological parameters, and so we vary the power spectrum of the initial conditions. Specifically, we define the primordial power spectrum as:
\begin{equation}
 P(k) = A_P \left(\frac{k}{ 0.78 \,\mathrm{Mpc}^{-1}}\right)^{n_P-1}\,.
 \label{eq:pk}
\end{equation}
Here $n_P$ is the spectral index measured on small scales. $A_p$ is the amplitude of perturbations at $k = 0.78 \,\mathrm{Mpc}^{-1}$. Note that $n_P$ and $A_P$ are distinct from $n_s$ and $A_s$, which are measured at $k_0 = 0.05$ Mpc$^{-1}$. We choose the scale $k = 0.78\, \mathrm{Mpc}^{-1}$ to reduce correlation between the amplitude and slope of the power spectrum measured by the forest \cite{Bird:2019}. We vary $A_P$ in the range $1.2 \times 10^{-9}$ and $2.6 \times 10^{-9}$, and $n_P$  between $0.8$ and $0.995$. These ranges are chosen to include the posterior constraint from Planck \cite{Planck:2018}.

We also include free parameters for the growth rate, $\Omega_M h^2$, and Hubble parameter, $h$. We vary $h$ over the range $0.65$ - $0.75$ and $\Omega_M h^2$ between $0.14$ and $0.146$. The growth rate $\Omega_M h^2$ has a moderate effect on the \Lya~forest, but is generally measured better by other probes. As our simulation box is set up using kpc/h units, the Hubble parameter $h$ does not affect the gravitational evolution of the code.

The cooling rates are computed in physical units, and so in principle the thermal history parameters may be correlated with $h$. However, as we will show in Section~\ref{sec:singleparams}, for the range of $h$ that we consider, the flux power spectrum is changed by only $1\%$. However, a subtle effect of cosmic variance means it is still important to simulate different values of $h$. More directly, changing $h$ at fixed $\Omega_M h^2$ changes $\Omega_M$. The flux power spectrum is measured in velocity units of s/km. The conversion between comoving kpc/h and km/s is
\begin{equation}
 1 \,\mathrm{km/s} =  100\, a \,(\Omega_M /a^3 + \Omega_\Lambda )^{1/2}\; \mathrm{kpc/h}\,.
\end{equation}
Our simulation boxes have a fixed size in kpc/h, so changing $\Omega_M$ moves the Fourier modes in the simulation box as measured in km/s. Sample variance in the simulation box scatters each Fourier mode around the true mean at fixed kpc/h. Thus when building an emulator on our simulation volume, some of the impact of sample variance is absorbed into a scale-dependent shift in the flux power spectrum and ascribed to $h$.

We fix $\Omega_b h^2 = 0.0224$ and vary $h$, so
the mass of a single gas particle varies between $5.3\times10^6$ and $7.0 \times 10^6 M_\odot h^{-1}$ for the low fidelity simulations and $6.7\times10^5$ and $8.0 \times 10^5 M_\odot h^{-1}$ for the high fidelity simulations. At linear order, $\Omega_b$ is mostly degenerate with the ionization fraction of the gas and thus the mean optical depth $\tau$, although it would change the strength of the BAO and potentially the effect of AGN feedback. However, both of these are relatively small effects and $\Omega_b h^2$ is well constrained by the CMB.

We do not include a parameter for massive neutrinos. Neutrinos are massless and included in the radiation, as the effect of massive neutrinos on these scales is completely degenerate with the amplitude of the initial perturbations, $A_P$ \cite{Pedersen:2020}.
Schematically, neutrino mass constraints come from comparing the power spectrum amplitude measured on CMB and \Lya~scales.

\subsection{Gravity, Hydrodynamics and Cooling}
\label{sec:gravity}

We use the \mpgadget~gravity solver, described in Ref.~\cite{Bird:2022} and references therein, with the hierarchical timestepping algorithm from Ref.~\cite{Springel:2021}. Long range gravitational forces are computed in Fourier space using a particle-mesh algorithm. Short-range forces, below the resolution of the grid, are computed using a hierarchical multipole expansion of the gravitational field, leading to a uniformly high force resolution throughout the computational volume. The default accuracy of the split between short and long range forces has been increased and the force accuracy is better than 1\% as measured against an exact N-body solver.

We adopt the pressure-entropy formulation of smoothed particle hydrodynamics (pSPH) to solve the Euler equations \citep{Hopkins:2013,Read:2010}. The implementation is discussed in Ref.~\cite{Feng:2014}. 
We use a cubic polynomial for the SPH density kernel, rather than the quintic kernel from \astrid, as the reduced number of neighbours improves resolution in the \Lya~forest. This comes at the cost of increased noise in dense regions, but this does not affect the gas dynamics at \Lya~forest densities \cite{Bird:2013}.
Gas is allowed to cool radiatively following Ref.~\citep{Katz:1996}.
We have used the updated cooling coefficients summarised in Ref.~\cite{Bolton:2017}, the recombination rates from Ref.~\cite{Verner:1996} and the collisional ionization rates from Ref.~\cite{Voronov:1997}. Self-shielding of neutral hydrogen is included following the fitting function of Ref.~\cite{Rahmati:2013}.
The UV background used is the ``optically thin'' variant from Ref.~\cite{FG2020}. We do not use the variant which corrects for the average effect of partial reionization on the gas temperature as we instead include explicit models for patchy reionization.

\subsection{Hydrogen Reionization}
\label{sec:hydrogen}

We model patchy reionization with a spatially varying ultra-violet background using a semi-analytic method \cite{Battaglia:2013}. We pre-compute a reionization redshift, on a grid with cells $1$ Mpc/h, using FASTPM \cite{FASTPM}. The redshift of reionization, $z_{HI}$, is defined as the redshift at which $50\%$ of the simulation volume has reionized. In general, higher initial overdensities reionize earlier. For gas particles in a region which has not yet reionized, the photon background is set to zero and the gas remains neutral. Once the reionization redshift has passed, the particle's ionization and thermal states evolve in ionization equilibrium with the external UV background.

At the time of reionization, each gas particle has its temperature boosted to $15,000$ K, to account for the (subgrid) heating effect of ionization fronts. Our patchy reionization model thus naturally includes effects of reionization which persist to low redshift \citep{Montero:2019}.
We base our model on the radiative transfer simulations of Ref.~\citep{DAloisio:2019}, who found a boost of $\sim 20,000$ K. However, we found that a temperature boost of $20,000$ K and a reionization redshift of $7$ produced a simulation with higher IGM temperatures than observed at $z\sim 6$ \cite{Gaikwad:2020}. Our simulation is much lower resolution and has much longer timesteps, suggesting that we should expect a certain amount of subgrid cooling between the passage of a reionization front and any given particle in our simulation box becoming active. 

We vary the reionization redshift $z_{HI}$ over the range $6.5 - 8$. For the highest $z_{HI} = 8$, $5\%$ of the box is reionized by $z=9.5$ and $95\%$ by $z=7$. The corresponding 5 and 95 \% volume quantiles for $z_{HI} = 6.5$ are $z=8$ and $z=5.5$, and for $z_{HI} = 7.25$, $z=6.2$ and $z=8.6$. Our emulator thus includes models where reionization finishes at $z < 6$, as suggested by some recent quasar dark gap measurements \cite{Zhu:2022}. After reionization the homogeneous UV background is not sufficiently intense to preserve a thermal equilibrium and so the gas cools. Since the IGM temperature is only measured for $z < 5.8$ \cite{Gaikwad:2020}, $z_{HI}$ is observationally degenerate with the size of the reionization temperature boost. Our ultimate constraints on the reionization redshift should thus be understood to be conditional on a boost of $15,000$ K.



\subsection{Helium Reionization}
\label{sec:helium}

We include a model for spatially inhomogeneous helium reionization following Ref.~\cite{UptonSanderbeck:2020}. Reionized bubbles are placed around potential quasars, randomly chosen halos in the mass range $10^{13} \geq M_{\rm halo}\geq 10^{12}$M$_{\odot}$. Bubbles are $20$ comoving Mpc/h ($\approx 25 - 30$ Mpc) in radius, matching the mean bubble size found in the radiative transfer simulations of Ref.~\citep{McQuinn:2009}. Inside a bubble a quasar radiation field is assumed with an effective spectral index of $ - \alpha_q$, which we vary in our simulation suite over the range $\alpha_q  = 1.3 - 2.5$. The quasar spectral index controls the level of heating during helium reionization and thus the peak IGM temperature. Higher values of $\alpha_q$ lead to lower peak temperatures.

Reionizing gas particles are abruptly heated and marked as ionized. Bubbles are placed until the total ionized gas mass in the box reaches
a pre-computed ionization fraction. We assume a linear ionization history between the initial redshift $z_{Hei}$ and the final redshift $z_{Hef}$. Following measurements of the optical depth in the Helium \Lya~forest \cite{Worseck:2019}, we simulate in the range $z_{Hei} = 3.5 -  4.1$ and $z_{Hef} = 2.6 - 3.2$. Our three free parameters allow us to generate a wide range of thermal histories, as shown in Figure~\ref{fig:meanigmtempdens}.

We checked that our results are not sensitive to the exact morphology of the reionization bubbles.
Specifically, we checked the effect of increasing the variance in the bubble size by running simulations ($30$ Mpc/h box side length, $256^3$ particles) with a mean quasar bubble size of $10$ Mpc with variances of $5$ Mpc and zero.
Setting the bubble size variance to zero decreased the temperature of the IGM by $\approx 1000$ K at redshifts $z<3$, which has the same affect as raising $\alpha_q$. This is likely a substantially larger effect than would be seen in the full box.

\begin{table*}
\begin{centering}
  \begin{tabular}{llll}
  \hline
  Parameter & Minimum & Maximum & Description \\
    \hline
    $n_P$  &  $0.8$  & $0.995$ & Scalar spectral index \\
    $A_P$  &  $2.2 \times 10^{-9}$  & $2.6 \times 10^{-9}$ & Power amplitude at $k = 0.78$ Mpc$^{-1}$ \\
    $h$    & $0.65$  & $0.75$ & Hubble parameter \\
    $\Omega_M h^2$ & $0.14$ & $0.146$ & Total matter density \\
    $z_{Hei}$      & $3.5$  & $4.1$  & Start redshift of HeII reionization \\
    $z_{Hef}$      & $2.6$  & $3.2$  & End redshift of HeII reionization \\
    $\alpha_q$     & $1.3$  & $2.5$ & Quasar spectral index during HeII reionization  \\
    $z_{Hi}$        & $6.5$ & $8$   & Median redshift of HI reionization \\
    $\epsilon_{AGN}$ & $0.03$ & $0.07$ & Thermal efficiency of black hole feedback \\
    \hline
  \end{tabular}
  \caption{Summary of varied emulator parameters, together with the ranges covered by the emulator. We vary a total of $9$ parameters: $4$ for cosmology, $3$ for the helium reionization model, $1$ for the hydrogen reionization model and $1$ for the strength of AGN feedback. }
  \label{tab:emulatorparams}
  \end{centering}
\end{table*}

\subsection{Star Formation and Stellar Feedback}
\label{sec:stellar}

Our star formation and stellar feedback model follows the \astrid~simulation and is explained in detail in Ref.~\cite{Bird:2022}.
Stars form on an effective equation of state in dense gas, following Ref.~\cite{Springel:2003}. Gas particles are converted directly to stars of the same mass. Note that \astrid~formed stars with $1/4$ of the mass of the gas particle. This increases the effective resolution of the stellar component of galaxies, but has no effect on the \Lya~forest, and so for our simulation suite simply increases computational cost. We disable metal return from stars, as it can be computationally expensive. Ref.~\cite{Viel:2013} showed that metal cooling has a small effect on the \Lya~forest flux power spectrum. Similarly, we do not include the (metallicity dependent) correction to the high redshift star formation rate due to the formation of molecular hydrogen implemented in \astrid, which has a very small effect for $z < 6$.\footnote{We thus cannot model the contamination of the \Lya~forest with metals from first principles with this suite alone, but as they are expected to be independent of cosmology we could do so using a separate simulation \cite[e.g. ASTRID~][]{Ni:2021, Bird:2022}.} 

A stellar wind feedback model is included following Ref.~\citep{Okamoto:2010} and Ref.~\cite{Bird:2022}. Wind speeds are proportional to the local one dimensional dark matter velocity dispersion $\sigma_\mathrm{DM}$:
\begin{equation}
v_w = \kappa_w \sigma_\mathrm{DM} \,,
\end{equation}
where $v_w$ is the wind speed. $\kappa_w$ is a dimensionless parameter, which we take to be $3.7$ following Ref.~\cite{Vogelsberger:2013}. In order to improve convergence with resolution, we added a minimum wind velocity, $v_w \geq 100$ km/s, following the IllustrisTNG model \cite{Pillepich:2018}.

Winds are sourced by newly formed star particles, which randomly pick gas particles from within their SPH smoothing length to become wind particles. The total mass loading is $(v_w/ 350 \mathrm{km/s})^{-2}$ where $350$ km/s is in physical units. Particles recouple when their surrounding density drops by a factor of $10$, or after $60$ Myr.\footnote{In Ref.~\cite{Bird:2022} we had a subdominant recoupling condition: gas would recouple after $20 \mathrm{kpc} / v_w$. In practice this affected only a small fraction of the wind, as for a typical star forming halo with a virial velocity of $200$ km/s the recoupling time was $100$ Myr, and gas always recoupled after $60$ Myr. We have thus removed this condition for model simplicity.} Particles in the wind cool, but do not experience or produce pressure forces, nor may they be accreted onto a black hole. However, to improve the stability of the SPH density estimates, they are included when computing SPH smoothing lengths.

We generated a small test suite, in $25$ Mpc/h boxes, where the dimensionless supernova wind velocity, $\kappa_w$, was varied. However, we found that this free parameter was essentially unconstrained by the \Lya~forest data and so opted not to vary it in the full emulator run. Ref.~\cite{Bolton:2017} showed that supernova winds increased the \Lya~flux power on large scales, $k < 10^{-2}$ s/km, at $z < 3$, by around $10\%$, due to the presence of additional high column density systems\footnote{Ref.~\cite{Viel:2013} found that the \Lya~flux power spectrum was increased only for $ k > 0.04$ s/km (scales smaller than those measured by BOSS), but their simulations did not include self-shielding and so did not contain high column density absorbers.}. Our supernova simulation model parameters have been chosen to match the observed galaxy stellar mass function and thus the distribution of high column density absorbers. We remove high column density systems from our simulated spectra to match the observational procedure, and thus changes to the supernova wind model do not affect the \Lya~flux power spectrum.

\subsection{Black Holes and AGN Feedback}
\label{sec:agn}

Our simulations include super-massive black hole (SMBH) seeding, growth and feedback following Ref.~\cite{Ni:2022}, itself based on \cite{Feng:2016,SDH2005,DSH2005}. Compared to \astrid, we have altered some of the thresholds to fit the lower resolution of our $1536^3$ simulations. SMBH particles are seeded by converting the densest gas particle found in a halo to a black hole. To be eligible for seeding, a halo must have total mass greater than $5\times 10^{10}\,h^{-1}M_\odot$ and stellar mass greater than $2 \times 10^9 h^{-1} M_\odot$. The SMBH seed mass is $5 \times 10^{5} h^{-1} M_\odot$. In the low fidelity simulations, the initial dynamic mass of the SMBH, which keeps the dynamical friction model stable for BH seeds, is set as $10^{8} h^{-1} M_\odot$. For comparison, the gas particle mass is $\sim 5 \times 10^6 h^{-1} M_\odot$ and the CDM mass is $ 3 \times 10^7 h^{-1} M_\odot$. SMBH are thus seeded only in well-resolved halos with (on average) $> 1400$ CDM particles and $>400$ star particles.

We implement a model for dynamical friction following Ref.~\citep{Chen:2021}. This ensures that SMBHs are kept close to the centers of their halos, and replaces the earlier manual repositioning. Dynamical friction is an artificial force modelling unresolved small-scale interactions between the SMBH and nearby stars. These interactions transfer momentum from the SMBH to individual stars in the surrounding star clusters. We include dynamical friction from stellar and CDM particles, but in practice the star particles strongly dominate \cite{Chen:2021}.

BHs are allowed to grow by accreting mass from nearby gas particles. Gas accretes following the Bondi-Hoyle-Lyttleton formula, applied to the smoothed properties of the gas particles within the SPH kernel of the BH:
\begin{equation}
\label{equation:Bondi}
    \dot{M}_{\rm B} = \frac{4 \pi \alpha G^2 M_{\rm BH}^2 \rho}{(c^2_s+v_{\rm rel}^2)^{3/2}}\,.
\end{equation}
$c_s$ and $\rho$ are the local sound speed and density of gas, $v_{\rm rel}$ is the relative velocity of the BH with respect to the nearby gas and $\alpha = 100$ is a dimensionless fudge parameter to account for the underestimation of the accretion rate due to the unresolved cold and hot phase of the subgrid interstellar medium nearby.
Note that hydrodynamically decoupled wind particles are not included in the density calculation of Eq.~\ref{equation:Bondi}. The accretion rate is capped at $2$ times the Eddington accretion rate.

AGN thermal feedback energy is implemented by depositing a fraction, $\epsilon_{AGN}$, of the available radiative energy for the AGN into the gas. We assume an accretion disk with a mass-to-light conversion efficiency of $0.1$, so that the AGN thermal energy is \cite{Shakura:1973}
\begin{equation}
\label{equation:Lbol}
    E_\mathrm{AGN} = \frac{\epsilon_{AGN}}{10} \dot{M}_{\rm BH} c^2\,.
\end{equation}
At high redshift, $z > 2$, most AGN feedback models are not efficient at removing gas from a halo, as there are few SMBHs in the low accretion regime necessary to induce kinetic mode feedback.
However, the \Lya~forest is sensitive to the gas temperature, and may be affected by gas heating around halos. Our test $25$ Mpc/h box simulations indeed showed some impact on the \Lya~forest flux power spectrum at $z=2.2$. We thus vary $\epsilon_{AGN}$  from $0.03 - 0.07$, around the fiducial \astrid~value of $0.05$ \cite{Ni:2022}. Section~\ref{sec:agnresult} shows, however, that in the larger boxes of our simulation suite, variations in $\epsilon_{AGN}$ have little impact on the \Lya~forest flux power spectrum. The effect in our test suite was due to sample variance in the small boxes, and because the boxes were too small to model the large-scale interaction of AGN with their environment.


\subsection{Generation of Spectra}
\label{sec:spectra}

We generate artificial spectra and compute the flux power spectrum as described in \cite{Bird:2019}. Each simulation generates output snapshots evenly spaced every $\Delta z = 0.2$ between $z = 5.4$ and $z = 2.2$. We generate spectra with a pixel width of $10$ km/s, finer than either the BOSS or DESI spectrograph. Spectra are generated using ``fake\_spectra'' \footnote{\url{https://github.com/sbird/fake_spectra}} \cite{FSFE:2017}. Our algorithm for generating artificial spectra differs from some other spectral generation codes, which compute the neutral hydrogen density on a 1D grid of spectral pixels, and then convolve each spectral pixel with the Voigt profile \cite[e.g.~][]{Theuns:1998, Chabanier:2022}. Thus for a pixel at position $x(j)$ the density, velocity and temperature $\rho$, $v$ and $T$ are computed in each bin by summing particles with neutral hydrogen density $\rho_i$:
\begin{align}
 \rho(x(j)) &= \Sigma_i \rho_i \\
 v(x(j)) &= \Sigma_i v_i \rho_i / \rho(x(j)) \\
 T(x(j)) &= \Sigma_i T_i \rho_i / \rho(x(j))\,.
\end{align}
The optical depth is computed by convolution with a Voigt profile for all positions $k$:
\begin{align}
 \tau(x(j)) &\propto \Sigma_k \rho(x(k)) \mathcal{V}\left[\frac{(v(k) - v(j))^2}{b(k)^2}\right] \\
 b(j) &= \sqrt{\frac{2 k_B T(j)}{m_P}}\,.
\end{align}
where $b(j)$ is the Doppler width of the line and $\mathcal{V}$ is the Voigt profile.

We instead skip the intermediate step of creating a density profile and compute the optical depth from each particle directly. In effect, each SPH particle is treated as an individual absorber, and the total absorption profile is the sum of the absorption from all particles:
\begin{align}
 \tau(x(j)) &\propto \Sigma_i \rho_i \mathcal{V}\left[\frac{(v(x(j)) - v_i)^2}{b_i^2}\right]\,.
 \end{align}
This preserves the ionization and velocity gradients inside a spectral pixel more accurately and shows better convergence, especially for dense gas \cite{Bird:2015}.


We compute the power spectrum of the flux overdensity, $P_F(k)$, as:
\begin{align}
 \delta_F(x) &= \mathcal{F}(x) / \widebar{\mathcal{F}} - 1 \\
 P_F(k) &= < L^{-1} \hat{\delta}_F(k) \hat{\delta}_F^*(k) >\,.
\end{align}
Here $\mathcal{F}$ is the (dimensionless) flux, related to the optical depth as $\mathcal{F}(x) = \exp{(-\tau(x))}$. $L$ is the length of the sightline. $\widebar{\mathcal{F}}$ is the mean flux, taken over all sightlines. We also define the mean optical depth, $\tau_\mathrm{eff} = -\log \widebar{\mathcal{F}}$. The overall flux power spectrum $P_F(k) $ is an average of the 1D Fourier transform of the flux, $\hat{\delta}_F(k)$, along each sightline.

We generate a regular grid of $480^2$ sightlines for each simulation snapshot, with a mean separation of $250$ kpc/h. We verified explicitly that we are converged with the number of sightlines: a grid of $720^2$ sightlines produced a flux power spectrum differing by less than $0.1\%$.\footnote{Recall that the gas initial conditions are a glass, not a grid, so the sightlines do not naturally align with the gas particles.} We generate three grids, one through each axis of the box. This allows the flux power spectrum to sample all independent Fourier modes of the matter overdensity field and thus reduces the sample variance in the flux power spectrum by $3$ on scales where the structure growth is linear.

We mask DLAs using a procedure similar to that used in the observational pipeline of Ref.~\cite{Chabanier:2020}. We first identify spectra where the maximum pixel optical depth $\tau > 10^6$ (corresponding to a column density $\sim 10^{20}$ cm$^{-2}$). We mark the region around this maximum optical depth where $\tau > 0.25 + \tau_\mathrm{eff}$, chosen to match the threshold of Ref.~\cite{Chabanier:2020}, which masks until the DLA is $20\%$ of the absorption. Within the masked region we set $\tau = \tau_\mathrm{eff}$, so that the masked region has $\delta_F = 0$. We checked that our flux power spectra changed by $< 1\%$ when the size of the masked region was increased by a factor of two. DLA masking affects approximately $2\%$ of spectra, in agreement with \cite{Rogers:2019}.

\subsubsection{Mean Flux}

Our simulated spectra are rescaled for a desired mean optical depth, parametrising the uncertainty in the ultra-violet background.
Since the mean optical depth does not require extra simulations, we dramatically over-sample it, generating a dense grid of $10$ optical depth samples per redshift per cosmological simulation. As explained in Section~\ref{sec:gpemulator} and Ref.~\cite{Bird:2019}, we will create one Gaussian Process emulator for every redshift bin, each emulator taking the mean optical depth at that redshift as an input parameter. We are thus able to use very general parametrisations for the redshift evolution of the mean optical depth.

However, in practice we found it sufficient to specialise the mean flux model to a power law. The amplitude $\tau_0$ and slope $d\tau_0$ of this power law are free parameters, defined relative to the value of $\tau_\mathrm{eff}$ measured by Ref.~\cite{Kim:2007}:
\begin{align}
\tau_\mathrm{Kim}(z) &= 0.0023 \times (1 + z)^{3.65} \\
 \tau_\mathrm{eff}(z) &= \tau_0 \left(\frac{1+z}{4}\right)^{d\tau_0}  \tau_\mathrm{Kim}(z)
 \label{eq:meanflux}
\end{align}
Thus by construction $d\tau_0$ does not affect the $z=3$ flux power spectrum, and $\tau_0 = 1$ and $d\tau_0 = 0$ correspond to the best-fit power law of Ref.~\cite{Kim:2007}. These parameters are used to calculate the mean optical depth in each redshift bin, and this mean optical depth is passed to the underlying Gaussian process for prediction.

\begin{figure}
    \centering
    \includegraphics[width=0.75\textwidth]{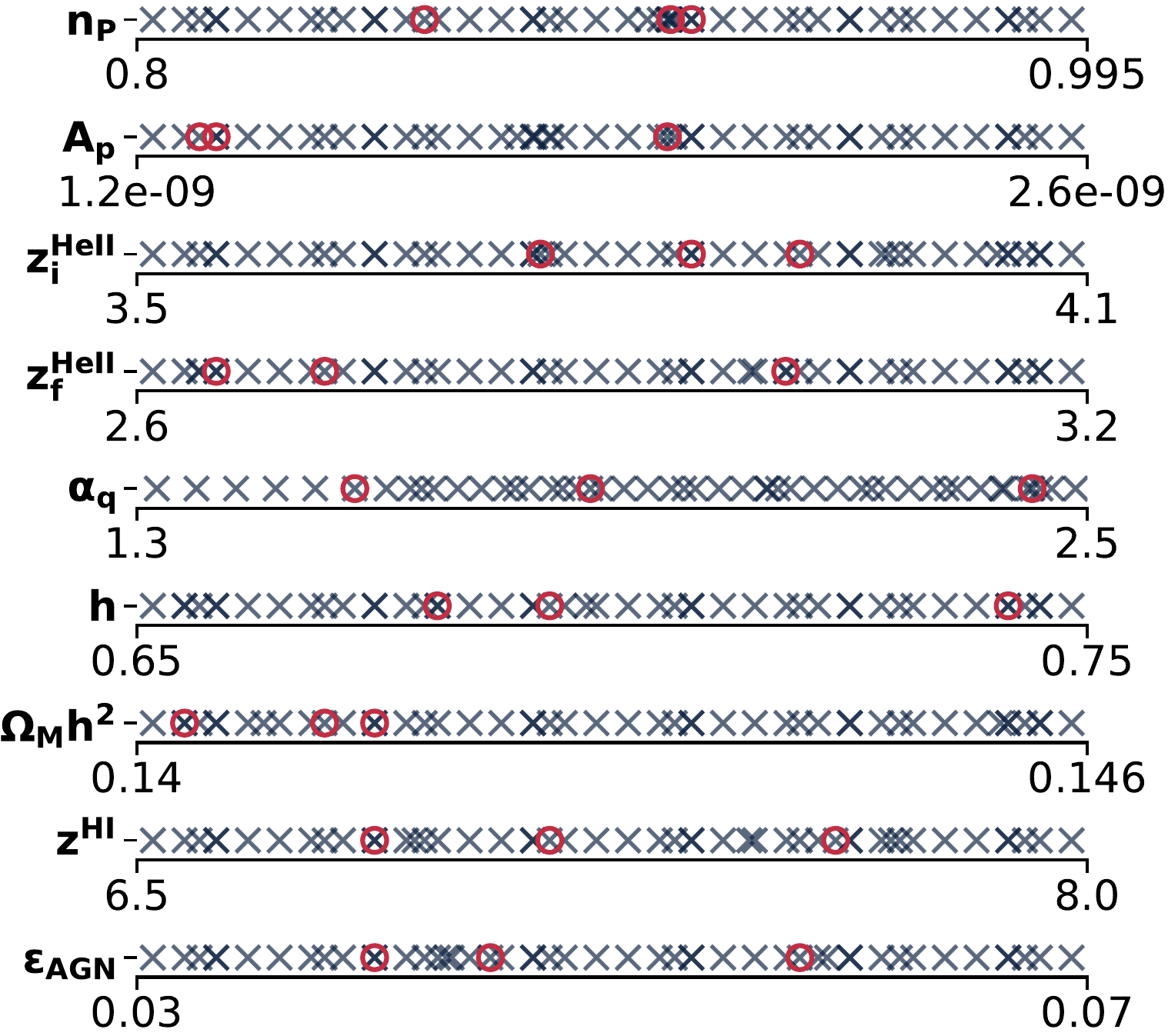}
    \caption{\label{fig:samples}
    Parameter limits for each of the nine parameters varied in the simulation suites.
    Grey crosses show samples at low fidelity. Red circles show samples at high fidelity.
    }
\end{figure}


\subsection{Experimental Design}
\label{sec:latinhypercube}


Our simulation suite includes a total of $10$ parameters at each redshift. One parameter, the mean flux, is generated in post-processing, meaning that we run simulations covering a $9$-dimensional parameter space, summarised in Table~\ref{tab:emulatorparams}.

We generate simulations at specific points in parameter space using a Latin Hypercube design, following the description in Ref.~\cite{Bird:2019}. Latin Hypercube samples are generated at random on a normalised unit cube, and the design which maximises the spread between parameter points is chosen for the final emulator. We ran $30$ simulations in our initial Latin Hypercube design. We then ran $2$ more low fidelity at parameters chosen by Bayesian optimisation \cite{Rogers:2019}.
However, we found that in our initial space Bayesian optimisation frequently chose points on the extreme boundaries of the parameter space.
We thus ran an extra $8$ simulation Latin Hypercube, before $3$ more low fidelity simulations whose simulation parameters were chosen with Bayesian Optimisation. Our initial suite covered $1.6 < \alpha_q < 2.5$. A comparison of our simulation outputs to the observed mean IGM temperatures suggested expanding the lower limit on this parameter, so we ran $6$ simulations uniformly spaced between $1.325 \leq \alpha_q \leq 1.575$. The parameters simulated are shown in Figure~\ref{fig:samples} (grey crosses), which demonstrates that both sets efficiently cover parameter space.

We then performed $3$ simulations at high fidelity, shown by the red circles in Figure~\ref{fig:samples}. The parameters to simulate were selected as suggested in Ref.~\cite{Ho:2022,Fernandez:2022}, by finding which $n$-simulation subset of the already complete low fidelity simulations produced the most accurate emulator. Specifically, the high fidelity samples were chosen by training a single-fidelity emulator using all combinations of two low fidelity samples to predict the remaining low fidelity samples for both the flux power and mean temperature \cite{Ho:2022}. The two low fidelity simulations which minimised the $L_2$-norm loss between the true and predicted flux power spectra and temperatures were then run at the higher resolution.

The third high fidelity sample was chosen similarly, but conditioned on the previously selected two samples. In practice, several potential simulations had similar losses, and so we chose the candidate with the lowest $A_p$, as these simulations have up to a factor of two lower computational cost.
Table~\ref{tab:highfidelity} shows the parameters of our $3$ high fidelity simulations.


\begin{table*}
\begin{centering}
  \begin{tabular}{llll}
  \hline
  Parameter & HF 1 & HF 2 & HF 3\\
    \hline
    $n_P$  &  $0.909$  & $0.914$ & $0.859$ \\
    $A_P$  &  $1.98 \times 10^{-9}$  & $1.32 \times 10^{-9}$ & $1.29 \times 10^{-9}$\\
    $h$    & $0.68$  & $0.74$ & $0.69$\\
    $\Omega_M h^2$ & $0.1403$ & $0.1415$ & $0.1412$\\
    $z_{Hei}$      & $3.75$  & $3.85$  & $ 3.92$ \\
    $z_{Hef}$      & $3.00$  & $2.65$  & $2.72$\\
    $\alpha_q$     & $2.43$  & $1.575$ & $ 1.87$ \\
    $z_{Hi}$        & $7.6$ & $6.875$   & $7.15 $\\
    $\epsilon_{AGN}$ & $0.045$ & $0.040$ & $ 0.058$\\
    \hline
  \end{tabular}
  \caption{Parameters of the three high fidelity (HF) simulations, performed with $2\times 3072^3$ particles.}
  \label{tab:highfidelity}
  \end{centering}
\end{table*}

\subsection{Multi-Fidelity Emulator}
\label{sec:gpemulator}

Having assembled a sample of simulated spectra, we use them to train a multi-fidelity Gaussian Process (GP) emulator for the flux power spectrum, following Refs.~\cite{Bird:2019,Ho:2022, Fernandez:2022}. Our emulator predicts the flux power spectrum for arbitrary cosmologies and redshifts within our input parameter range, along with a measure of prediction uncertainty. Our multi-fidelity emulator has two parts. First, a Gaussian process is built to predict the results of a low fidelity simulation at arbitrary cosmology. Then a second Gaussian process is used to model a cosmology dependent correction function between low fidelity and high fidelity simulations. Many low resolution simulations are used to explore parameter space and only a few high resolution evaluations are needed to understand the correction function.


We use the GPy package \cite{gpy2014} and EmuKit \cite{2021arXiv211013293P}. If $P_F(\boldsymbol{\theta})$ is the simulated \Lya~forest flux power spectrum as a function of a parameter vector $\boldsymbol{\theta}$, then a GP models this output as draws from a distribution
\begin{equation}
    P_F(\boldsymbol{\theta}) \sim GP(\mu(\boldsymbol{\theta}), k(\boldsymbol{\theta}, \boldsymbol{\theta}^{\prime})),
\end{equation}
where $\mu(\boldsymbol{\theta})$ and $k(\boldsymbol{\theta}, \boldsymbol{\theta}^{\prime})$ are the mean and covariance function, respectively.
As in Ref.~\cite{Fernandez:2022}, we rescale the training samples by the median of the low fidelity spectra, so that they match the GP prior of a zero mean function. The predictions are then multiplied by this rescaling factor on output. The covariance kernel is the combination of a radial basis function and a linear kernel. The total kernel is
\begin{equation}
        k_\mathrm{RBF}(\boldsymbol{\theta}, \boldsymbol{\theta}'; \sigma_0, \boldsymbol{l}) + k_\mathrm{LIN}(\boldsymbol{\theta}, \boldsymbol{\theta}'; \boldsymbol{\sigma})\\
        = \sigma_0^2 \exp{\left( \sum_{i=1}^{d} -\frac{(\boldsymbol{\theta}_i - {\boldsymbol{\theta}_i}')^2}{2 l_i^2} \right)} +  \sum_{i=1}^{d} \sigma_i^2 \boldsymbol{\theta}_i {\boldsymbol{\theta}_i}',
        \label{eq:kernel}
\end{equation}
where $d$ is the dimensionality of the input parameters.

The multi-fidelity part of the model uses the linear multi-fidelity model \citep{10.1093/biomet/87.1.1}. Using this model, which Ref.~\cite{Fernandez:2022} showed to be accurate, a high fidelity prediction is given by
\begin{equation}
    P_F^{^\mathrm{HF}}(k, z, \boldsymbol{\theta}) = \rho(k, z) \cdot P_F^{^\mathrm{LF}}(k, z, \boldsymbol{\theta}) + \delta(k, z, \boldsymbol{\theta}),
    \label{eq:ko_model}
\end{equation}
where $\rho$ is a scaling parameter, and $\delta(\boldsymbol{\theta})$ is a GP (independent of the LF output). Both $\rho$ and $\delta$ are optimized using the training samples. $\rho$ is a multiplicative correction, and $\delta$ is an additive correction. Notice that the cosmology dependence comes from $\delta$. In practice we shall see that the cosmology dependence of the resolution correction is fairly small for our simulations, and so the linear multi-fidelity model is a good choice. Redshift dependence is incorporated by building separate GP models for each redshift bin.

The hyperparameters that are learned from the training samples are: variances for the RBF ($\sigma_0^2$) and  linear ($\boldsymbol{\sigma}^2$) kernels, and the lengthscale controlling the smoothness of the RBF kernel, $\boldsymbol{l}$. An independent value is assigned for each dimension $d$ of the input for each of these hyperparameters. Parameters to which the flux power spectrum is more sensitive have a smaller scale. The kernel for $\delta(\boldsymbol{\theta})$ is also the combination of an RBF and a linear kernel. However, all parameters share a single length scale. This simplification was done to improve the training time, and we verified that it did not affect prediction accuracy. With only $3$ high fidelity samples, there is in any case little data to separately optimise the hyperparameters of $\delta(\boldsymbol{\theta})$.

Ref.~\cite{Fernandez:2022} trained a GP for every $k$ bin for every redshift. However, with our increased training set size, we found that training a GP for each $k$ bin led to very large memory usage for the emulator, and substantially slowed down the training and prediction steps. Here we instead train a single GP for each redshift bin, across the full $k$ range, inducing a correlation between outputs at different $k$-bins. We checked explicitly that this had a very small effect on our interpolation accuracy at all scales.  Our low fidelity simulations are already almost converged, making the correction function is easier to model and avoiding the need for separate hyperparameters in each $k$-bin.
The output of each GP is thus a vector of $P_F(k)$ for specific $k$ bins.


\subsubsection{Leave One Out Errors}
\label{sec:looerror}

\begin{figure}
    \centering
    \includegraphics[width=\textwidth]{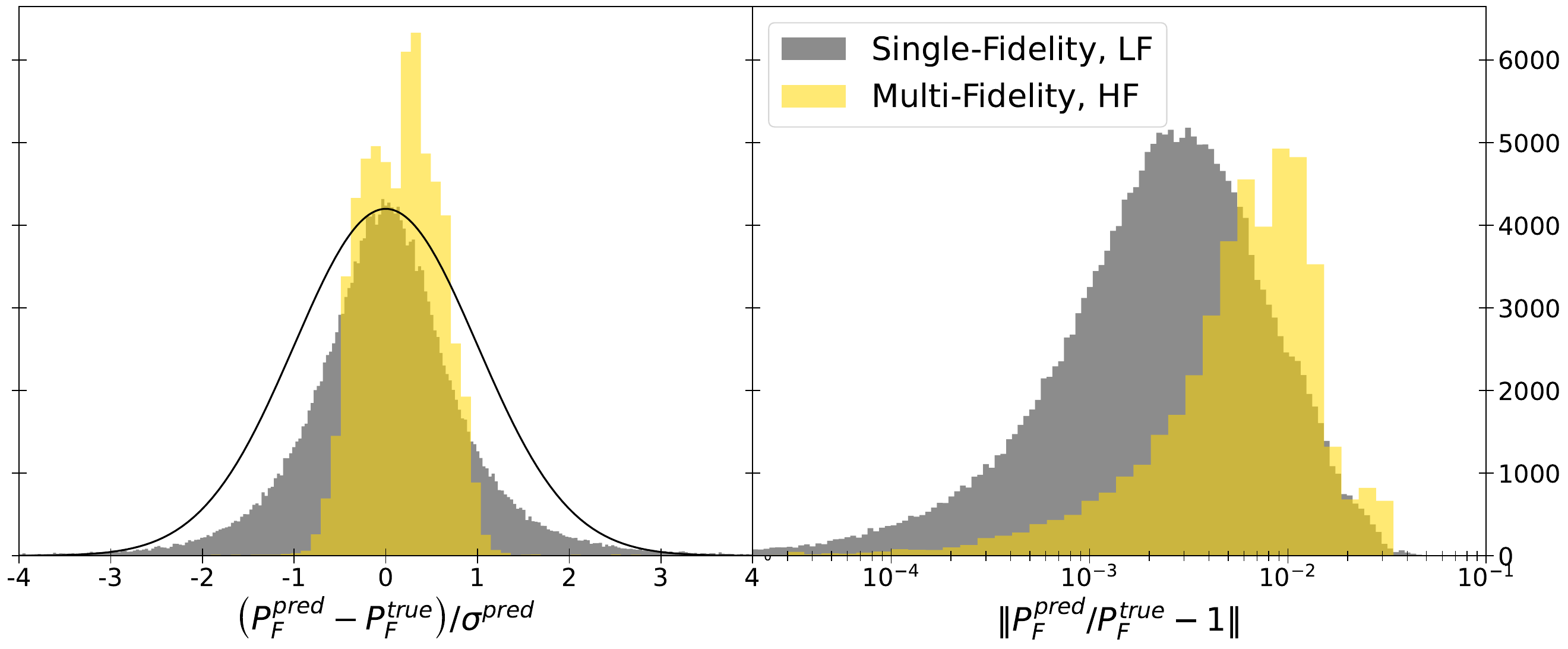}
    \caption{\label{fig:fps_error}
    Leave-one-out errors for the \Lya~forest flux power spectrum emulator.
    The left panel shows the error in units of prediction uncertainty. The solid black line shows a unit Gaussian normalised to the central height of the emulator error.
    The right panel shows the relative errors, an estimate of how well the emulator predicts the flux power spectrum.
    The histograms show the prediction error for each $k$-bin at each redshift of each flux power sample included in the final training sample.
    The grey distribution shows the emulator error for a single-fidelity emulator, predicting the flux power for each of the $48$ LF simulations from a $47$ simulation emulator.
    The yellow distribution shows the errors for a multi-fidelity emulator, predicting the flux power for each of the $3$ HF simulations from a $48$ LF, $2$ HF simulation emulator.
    }
\end{figure}

We measured the interpolation accuracy for our emulator using leave-one-out errors. Each simulation is left out of the emulator training set in turn. A new emulator is trained on this reduced set, and a prediction is made for the omitted simulation. All samples from the same simulation, even those with different values for the mean flux, are omitted simultaneously, as they are highly correlated. The relative error between the prediction and true output from the omitted simulation is then used as the leave-one-out score.
The distribution of relative errors from this calculation are shown for the flux power spectrum in Figure~\ref{fig:fps_error}. We display the scales probed by eBOSS, $10^{-3} < k_F < 2 \times 10^{-2}$ s/km.
We show the errors for a single-fidelity emulator that is predicting low fidelity outputs (grey), and the errors for a multi-fidelity emulator that is predicting high fidelity outputs (yellow). As there are only three high fidelity simulations, the multi-fidelity emulator error distributions have much smaller sample sizes, and have been scaled for clarity.

In the left panel of Figure~\ref{fig:fps_error}, the absolute error normalized by the prediction uncertainty is shown. This gives an indication of the calibration of our emulator errors. The black line shows a unit Gaussian normalised to the same peak value as the single fidelity error histogram. The interpolation errors are noticeably non-Gaussian, because we train a single emulator for all k bins, which creates significant correlations between scale bins. In addition, the leave-one-out errors are generally $\sim 50\%$ smaller than the errors predicted by the emulator.

The right panel shows the absolute relative error, a good indicator of the accuracy of the emulator predictions.
The low fidelity flux power emulator has an average error of $\sim 2\times 10^{-3}$. For the high fidelity predictions, the median error is larger, around $10^{-2}$. Note that leave-one-out errors by definition are computed using smaller emulation sets than the full emulator, and are thus conservative. The multi-fidelity leave-one-out errors in particular are computed using an emulator built with only $2/3$ high-fidelity simulations, and so missing $30\%$ of the information. The worst-case leave-one-out error predicting the low fidelity flux power spectrum was $\sim 0.06$. This prediction is for a simulation on the extreme edge of parameter space ($A_P = 2.57\times10^{-9}$), and so an extrapolation for the leave-one-out emulator.

Ref.~\cite{Fernandez:2022} achieved sub-$1\%$ accuracy using $30$ low fidelity and $5$ high fidelity simulations. We have achieved similar accuracy with only $3$ high fidelity simulations. We need a smaller set of high fidelity simulations because our low fidelity training set has a mean inter-particle spacing of $78$ kpc/h, whereas Ref.~\cite{Fernandez:2022} use low fidelity simulations with a mean inter-particle spacing of $117$ kpc/h. The low fidelity simulations are thus more nearly converged and need a smaller correction factor.

For comparison, the diagonal elements of the covariance function of the flux power spectrum from DR14 are between $1.5\%$ (at $z=2.8$ and k = $5\times 10^{-3}$ s/km) and $\sim 15\%$ (at $z=4.6$) \cite{Chabanier:2019}. Thus, although our interpolation errors are not perfectly calibrated, the conservative estimates above are smaller than the observational uncertainties.

\subsubsection{Mean IGM Temperature and Emulator}

\begin{figure}
    \centering
    \includegraphics[width=\textwidth]{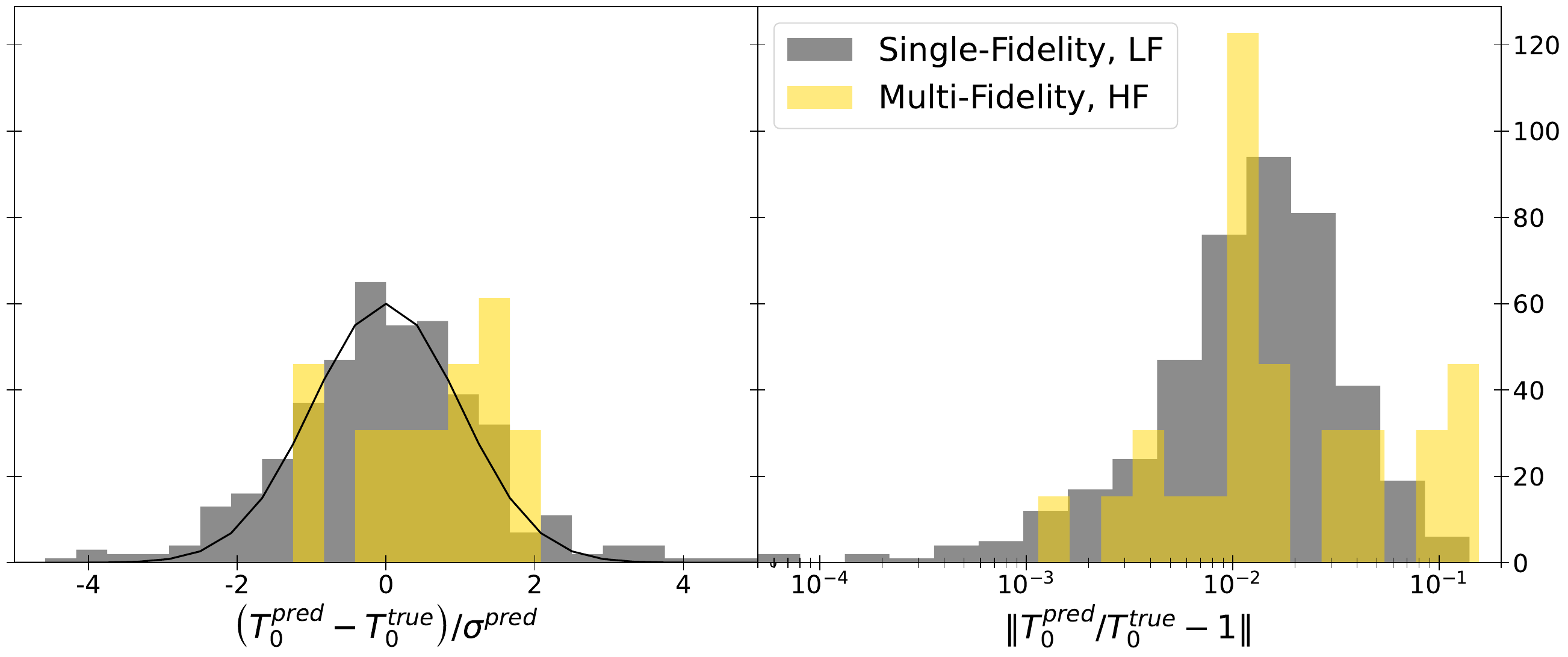}
    \caption{\label{fig:t0_error}
    Leave-one-out errors for the IGM mean temperature emulator.
    The left panel shows the error in units of prediction uncertainty. The solid black line shows a unit Gaussian normalised to the central height of the emulator error. The right panel shows the relative errors.
    The grey distribution shows the emulator error for a single-fidelity emulator, predicting the mean temperature for the LF simulations.
    The yellow distribution shows the errors for a multi-fidelity emulator, predicting the mean temperature for the HF simulations. Solid black line shows a unit Gaussian normalised to the central height of the emulator error.
    }
\end{figure}

We define the IGM mean temperature from the simulation snapshots as the median temperature for all particles within $5\%$ of the critical density.
The inherent patchiness of helium reionization in our model implies that the simulation box contains $\sim 50$ bubbles differing in temperature and pressure smoothing. A consequence is that there is no unique temperature-density relation in our simulations, but a range of temperature-density relations depending on the time at which a given part of the simulation reionized (see Ref.~\cite{UptonSanderbeck:2020}). We have extracted the mean IGM temperature from all simulation snapshots in our suite.

We then built an emulator for the mean IGM temperature, using the same kernel as for the flux power spectrum (Eq.~\ref{eq:kernel}), and the same linear multi-fidelity model (Eq.~\ref{eq:ko_model}). Figure~\ref{fig:t0_error} shows the leave-one-out errors for the mean IGM temperature emulator. The left panel shows the leave-one-out errors as a ratio to the expected emulator errors. These are a good match to a unit Gaussian, showing that the mean temperature emulator has well-calibrated errors. The right panel shows the absolute values of the leave-one-out emulator errors, demonstrating that the median error is $1\%$, with a worst-case error of about $10\%$, comfortably below current observational uncertainties \cite{Gaikwad:2020}.

\section{Results}
\begin{figure*}
\includegraphics[width=1.0\textwidth]{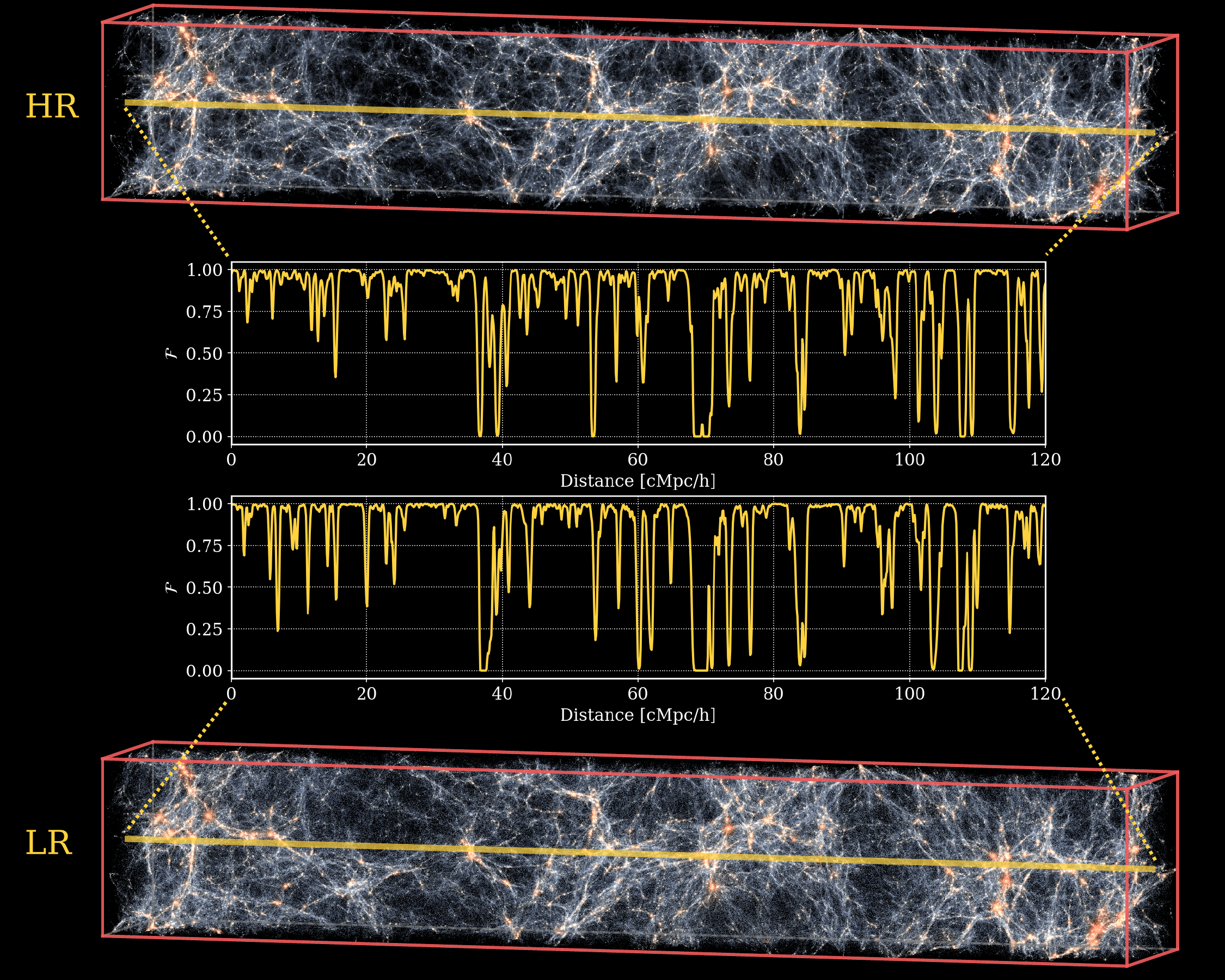}
 \caption{Visualisations of the HF1 (see Table~\protect\ref{tab:highfidelity}) low and high resolution simulations at $z=2.2$. Shown are $20\times 20$ Mpc/h tubes across the full $120$ Mpc/h box. A sightline is drawn through the center of each simulation box, and the spectra visualised. The colours indicate the density of the gas, weighted by internal energy, with redder colours indicating higher temperature. Low and high fidelity simulations are similar, demonstrating that our low fidelity simulations are already reasonably well converged.}
 \label{fig:visualisation}
\end{figure*}

In this Section we show some results from our simulation suite. Figure~\ref{fig:visualisation} shows a visualisation of the low and high fidelity simulations at $z=2.2$, through a region $20$ Mpc around the center of the box. We evaluate convergence with box size and resolution in Section~\ref{sec:boxsize}, showing that the high fidelity simulations are well converged, and visualising the correction function of the multi-fidelity emulator. In Section~\ref{sec:thermal} we show the thermal histories covered by our emulator, together with the convergence of the thermal history with resolution. Section~\ref{sec:singleparams} displays how the flux power spectrum and mean IGM temperature depend on the different parameters in our simulation suite. We specifically discuss the effect of AGN feedback on the flux power spectrum in Section~\ref{sec:agnresult}. Finally, we verify that our simulation suite produces a realistic population of DLAs in Section~\ref{sec:dlas}.

\subsection{Box Size and Particle Load}
\label{sec:boxsize}

\begin{figure*}
\includegraphics[width=0.95\textwidth,trim={0 4.5cm 1cm 0},clip]{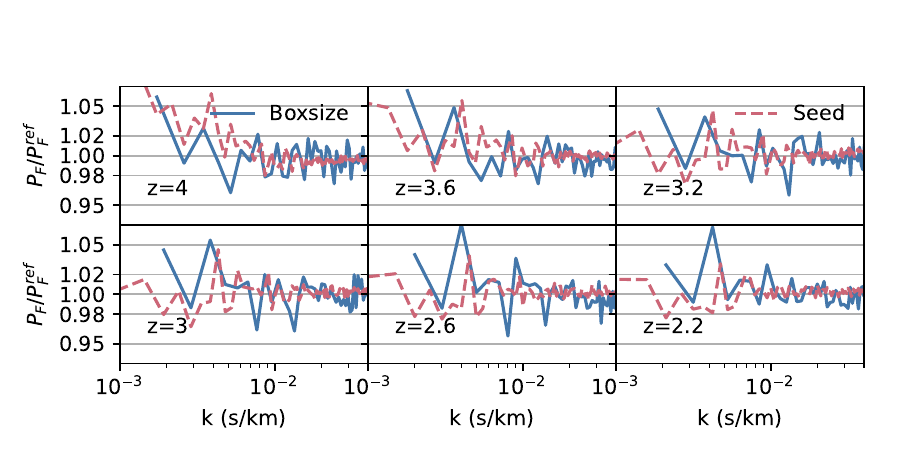}
 \caption{Convergence of the flux power spectrum with box size and sample variance. Blue solid lines (``Boxsize'') show the ratio of the flux power spectrum as a function of redshift in a $120$ Mpc/h box compared to a $60$ Mpc/h box. Our primary simulations use a $120$ Mpc/h box. This is thus an upper limit on the convergence with respect to the simulation box. Red dashed lines (``Seed'') show the ratio between one of our low fidelity simulations in a $120$ Mpc/h and a separate simulation with a different initial structure seed, also in a $120$ Mpc/h box, demonstrating the impact of sample variance. The range of $k$ measured by eBOSS is $1.1 \times 10^{-3}$ s/km to $2\times 10^{-2}$ s/km.}
 \label{fig:boxsize}
\end{figure*}

Figure~\ref{fig:boxsize} shows the effect of the finite box size on the flux power spectrum, as a function of redshift. We show the flux power spectrum ratio between a test simulation in a $60$ Mpc/h simulation box and a larger $120$ Mpc/h simulation box. The mean inter-particle spacing is $120$ kpc/h for both simulations, so that the large box simulation has $1024^3$ particles, while the small box simulation has $512^3$ particles. Since our main simulations use $120$ Mpc/h simulation boxes, this is an upper limit on the effect of the finite box. We checked that the largest scale at which non-linear growth changes the matter power spectrum by $\sim 10\%$ at $z=2$ is $\sim 45 - 60$ Mpc/h. Box size convergence is at the level of $2\%$, with the exception of two modes near the edge of the $60$ Mpc/h box, and appears dominated by cosmic variance in the smaller box.

Figure~\ref{fig:boxsize} also shows the effect of cosmic variance in our simulation suite, by finding the ratio of two simulations with different initial structure seeds. For $z < 3.6$ cosmic variance is important at the level of $2\%$. Cosmic variance is suppressed when compared to emulators of the matter power spectrum as our summary statistic is the $1$-D flux power spectrum, evaluated along the line of sight to the quasar and averaged over multiple quasar positions. Effectively this sums over two dimensions of the box, ensuring that even the largest scales sample Fourier modes proportional to the number of spectra. There is a $\sim 2-5\%$ effect from cosmic variance for $k < 3 \times 10^{-3}$ s/km at $z \geq 4$. On these scales the total error budget from BOSS DR14 is $> 10\%$, so cosmic variance is subdominant to the observational errors. However, it is curious that cosmic variance is important at higher redshifts, when the structure might be expected to be more linear. A plausible explanation is that different structure seeds have a different reionization history, which in turn affects the flux power spectrum.

\begin{figure*}
\includegraphics[width=0.95\textwidth,trim={0 0 1cm 0},clip]{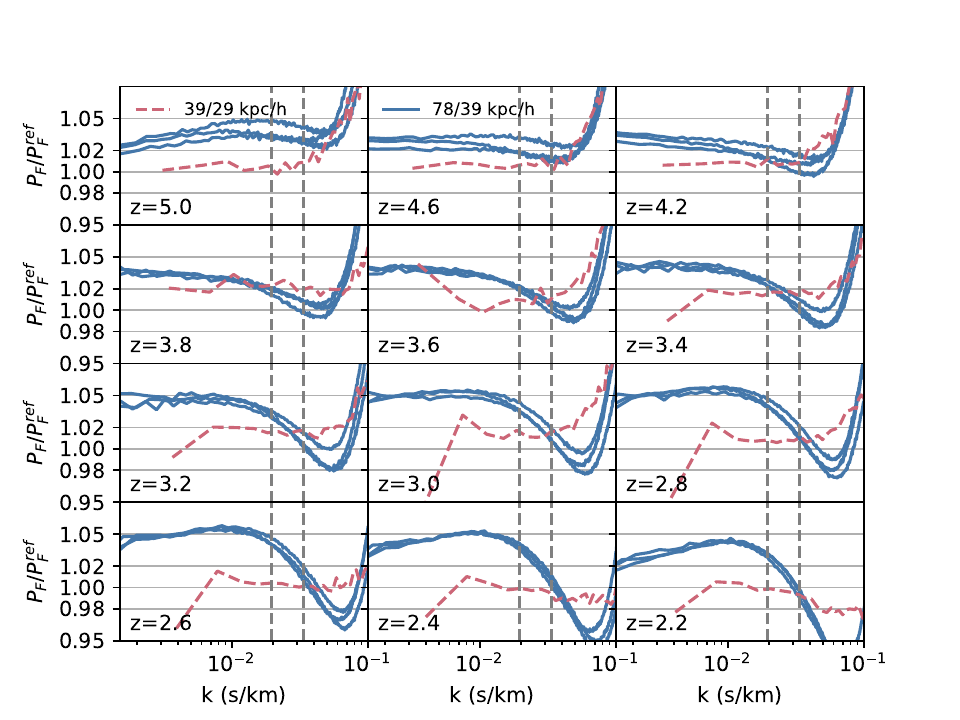}
 \caption{Convergence of the flux power spectrum with resolution (mean interparticle spacing). Each line labelled $78/39$ kpc/h (blue solid) compares a single LF/HF simulation pair. One line is shown for each of the $3$ HF simulations. The LF simulation in each pair has $1536^3$ particles in a $120$ Mpc/h box, and thus a mean-interparticle spacing of $78$ kpc/h, while the HF simulation has $3072^3$ particles and thus $39$ kpc/h. The line labelled $39/29$ kpc/h compares a pair of simulations in a $15$ Mpc/h box. The first has $384^3$ particles, and the second has $512^3$ particles. The figure shows that our HF simulations are converged, but that the multi-fidelity emulator is significantly correcting the LF simulations.}
 \label{fig:resolution}
\end{figure*}

Figure~\ref{fig:resolution} shows the effect of finite resolution on the flux power spectra outputs of our simulations. We show in blue the difference between the high and low resolution branches of the simulation suite, for the three high fidelity simulations. Note that these simulations have been rescaled to have the same mean optical depth, as these are the values input into the emulator. Our lower resolution simulations contain $2\times 1536^3$ particles in a $120$ Mpc/h periodic box with a mean inter-particle spacing of $79$ kpc/h. Our $3$ higher resolution simulations contain $2\times 3072^3$ particles and thus a mean inter-particle spacing of $39$ kpc/h. Figure~\ref{fig:resolution} essentially shows the resolution correction function of our multi-fidelity emulator, as described in Eq.~\ref{eq:ko_model}. The dependence of the resolution correction on cosmological parameters is fairly small, so the prediction is dominated by the cosmology-independent scaling factor $\rho$ in Eq.~\ref{eq:ko_model}. Our high fidelity node is meaningfully correcting the resolution of our low fidelity simulations, as $\rho$ is not unity.
However, the convergence in our low fidelity node is already reasonably good, at the level of $5\%$ for $k < 0.05$ s/km, the scales measured by DESI (and eBOSS).

Ref.~\cite{Fernandez:2022} required six high fidelity simulations to achieve good interpolation accuracy. The low fidelity node of the emulator constructed in Ref.~\cite{Fernandez:2022} had a mean interparticle spacing of $118$ kpc/h\footnote{We originally intended this to be the resolution of the low resolution node of our current emulator, but recent improvements in \mpgadget~allowed us to increase our minimum particle load.}. Our current low fidelity simulations have a mean interparticle spacing of $78$ kpc/h, $1.5$ times smaller, which allows us to reach high interpolation accuracy with only $3$ high fidelity simulations.

To demonstrate convergence of our high fidelity node we also show a comparison between two $15$ Mpc/h box size simulations, one $384^3$ particles and thus the same resolution as our high fidelity node, and one with $512^3$ particles. Convergence is at the $2\%$ level for $z > 2.6$ and the sub-percent level for $z \leq 2.6$, apart from scales greater than $1/4$ the box size. The simulation has $z_{HI} = 7.7$, and helium reionization goes from $z=4 - 3.1$, with $\alpha_q = 2.37$. This level of resolution sensitivity is small compared to current observational uncertainties.
The worst convergence is seen during helium reionization, which starts at $z=4$ and ends at $z=3.1$. This may indicate that the simulation is being affected by the small separation between the size of the box and the size of the helium reionizing bubbles, and that a hypothetical very high resolution simulation in a larger box would display better resolution convergence. In addition, the small box will lose power at low redshifts, which is likely the source of the poorer convergence at $k < 0.01$ s/km.

The scales measured by high resolution \Lya~forest spectra (XQ-100 and KODIAQ/SQUAD) are $ k < 0.1$ s/km, with a $\sim 10\%$ statistical error \cite{Irsic:2017, 2022MNRAS.509.2842K}. The approximate limit of the $2\%$ convergence for our high fidelity simulations is $0.07$ s/km, and we are converged to about $7\%$ at $0.1$ s/km. A comparison to high resolution data using our simulations is thus possible, and will be addressed in future work.

\subsubsection{Comparison to Earlier Convergence Results}
\label{sec:reionresolution}

Our resolution convergence is better than expected from earlier simulation suites (although even with convergence behaviour or other simulations our resolution would still be sufficient to model the scales and redshifts observed by eBOSS and DESI). Ref.~\cite{Borde:2014} performed a convergence study using a $20$ Mpc/h box and $192^3$ - $1024^3$ particles. Our high fidelity node is equivalent in resolution to their $20$ Mpc/h, $512^3$ simulation, and exhibits similar convergence for $z < 3.6$, although our simulations are fully hydrodynamic while theirs make use of the simplified \Lya~forest star formation model. However, at $z=4.4$, their simulation loses $2-5\%$ of the flux power for $k > 0.01$ s/km. Although this degree of convergence is negligible compared to the observational errors at $z=4.4$, we do not observe a similar effect.
We also found that our convergence at $z \sim 5$ was better than predicted by Ref.~\cite{2009MNRAS.398L..26B, Boera:2019}. This is primarily due to our inclusion of a temperature boost during hydrogen reionization \cite{DAloisio:2019}. The temperature boost heats the gas, increasing the thermal free-streaming length and pressure smoothing scale \cite{Bolton:2017}. This prevents small overdensities, which would normally be under-resolved, from accreting gas. Note that the majority of reionization happens for $z > 6$ in all our reionization
models, so we do not expect the convergence to be sensitive to the model parameters. Even with the same reionization model, we would expect our simulations to be slightly more converged than earlier work as we have increased the accuracy parameters for MP-Gadget's gravitational integration \cite{Bird:2020}, and our mean optical depth assignment scheme appears to exhibit better convergence than some earlier spectral estimation techniques \cite{2015MNRAS.447.1834B}.


\subsection{Thermal Histories}
\label{sec:thermal}

\begin{figure*}
\includegraphics[width=0.45\textwidth]{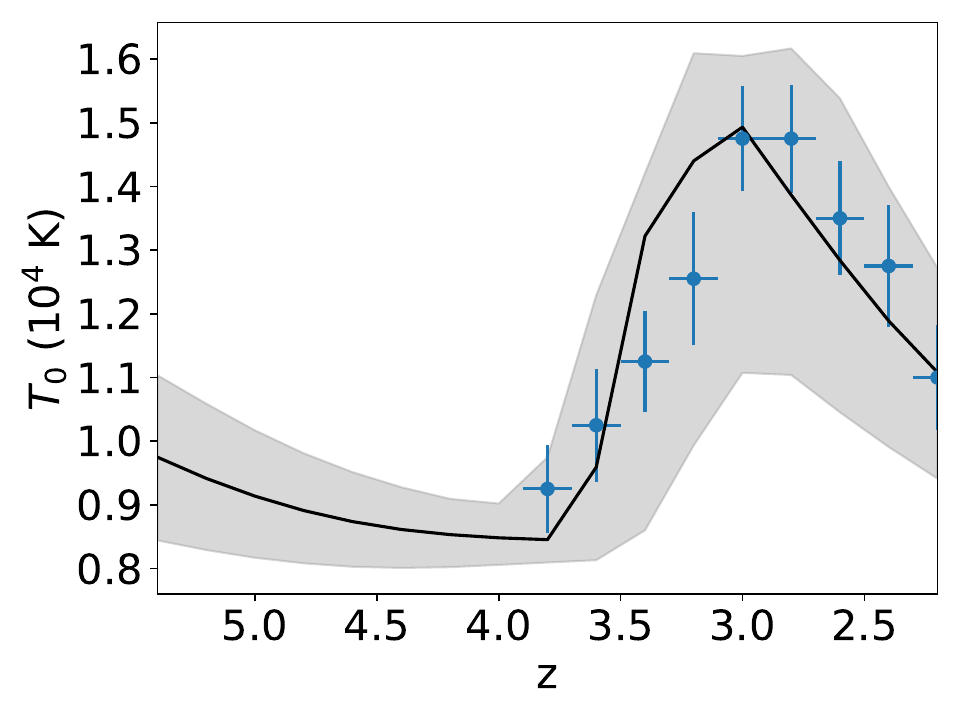}
\includegraphics[width=0.45\textwidth]{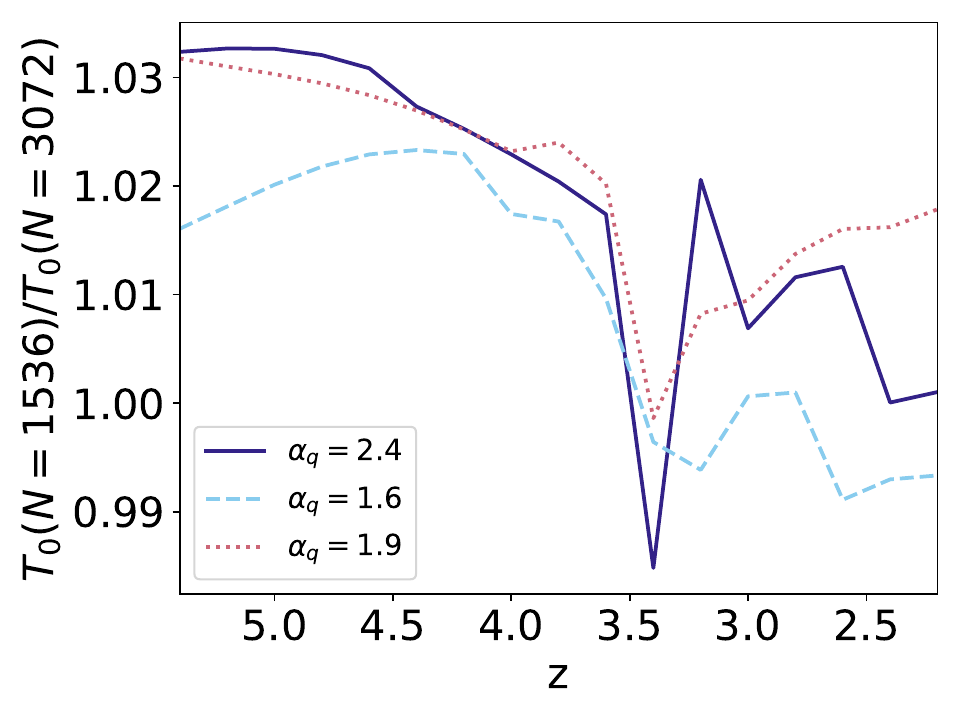}
 \caption{(Left) Temperatures at mean density as a function of redshift for simulations in our emulator. Observational points shown are the combined $T_0$ measurements from \protect\cite{Gaikwad:2021}.  The grey band shows the range of temperatures realised by simulations in our emulator. The black line shows an example thermal history, for a simulation with reionization parameters $\alpha_q = 1.80$, $z_{Hei} = 3.79$, $z_{Hef} = 3.09$, $z_{Hi} = 7.08$. Other parameters are $n_P = 0.823$, $A_P = 1.92 \times 10^{-9}$, $\Omega_M h^2 = 0.1445$, $h = 0.69$, $\epsilon_{AGN} = 0.067$.
 (Right) Ratio between the IGM mean temperature in low resolution simulations to the high resolution simulations, each labelled by their $\alpha_q$ value. Convergence is at the level of $3\%$ while the observational error on the temperature is $5-10\%$.
}
 \label{fig:meanigmtempdens}
\end{figure*}

Figure~\ref{fig:meanigmtempdens} shows the mean temperatures for our simulations, confirming that our reionization models are able to bracket the observed mean IGM temperatures from Ref.~\cite{Gaikwad:2021}.
The right panel shows the convergence with resolution, for our three high fidelity simulations, labelled by their $\alpha_q$ values. Convergence is better than $3\%$ between the low and high fidelity simulations, showing that even our low fidelity nodes contain converged thermal histories.
The high fidelity simulations generally have slightly lower temperature than the low fidelity simulations, as the inclusion of higher density structures increases the amount by which gas can cool. The high fidelity simulation with the lowest $\alpha_q$ value has the strongest heating during helium reionization (as well as the lowest redshift of hydrogen reionization). The extra heating reduces the amount of dense gas and thus improves the overall convergence of the simulation.



\subsection{Effect of Parameters on the Flux Power Spectrum and Temperature}
\label{sec:singleparams}

\begin{figure}
    \centering
    \includegraphics[width=0.48\columnwidth]{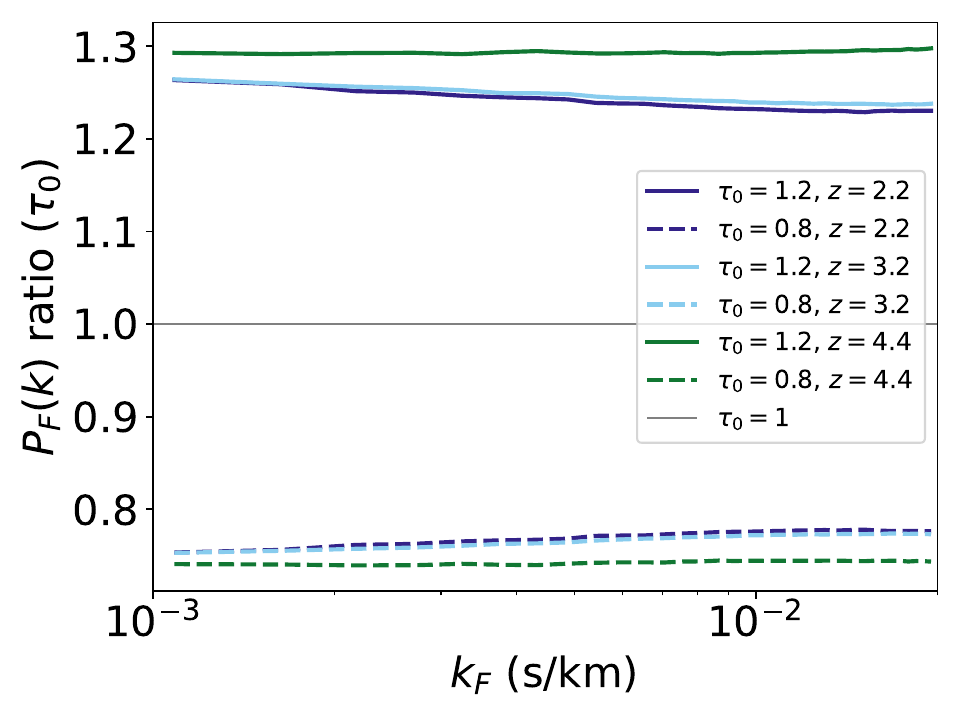}
    \includegraphics[width=0.48\columnwidth]{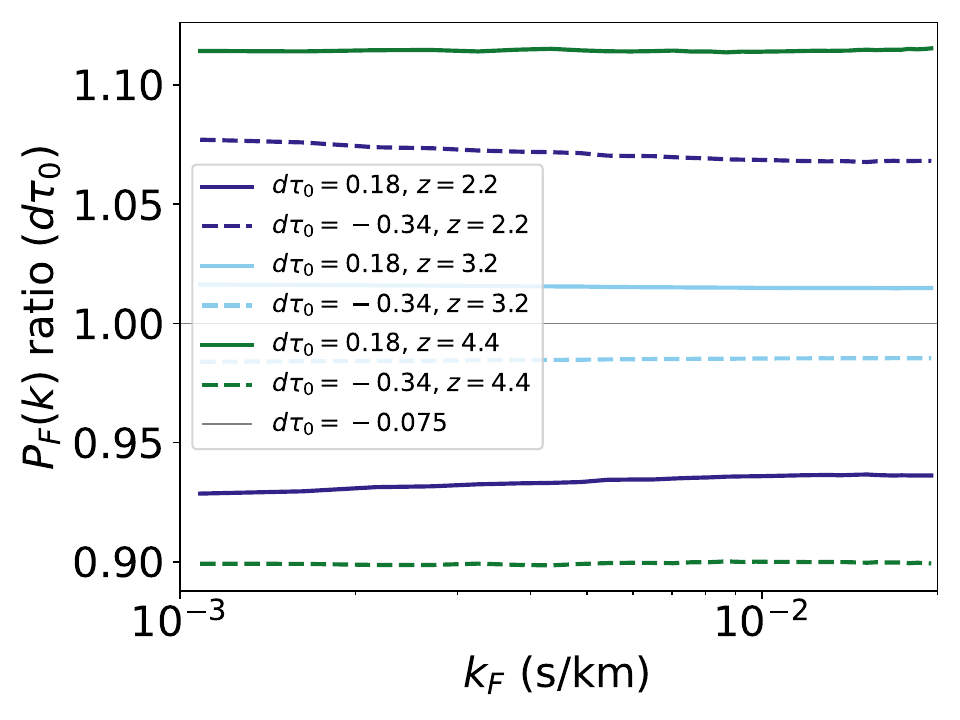}
	\includegraphics[width=0.48\columnwidth]{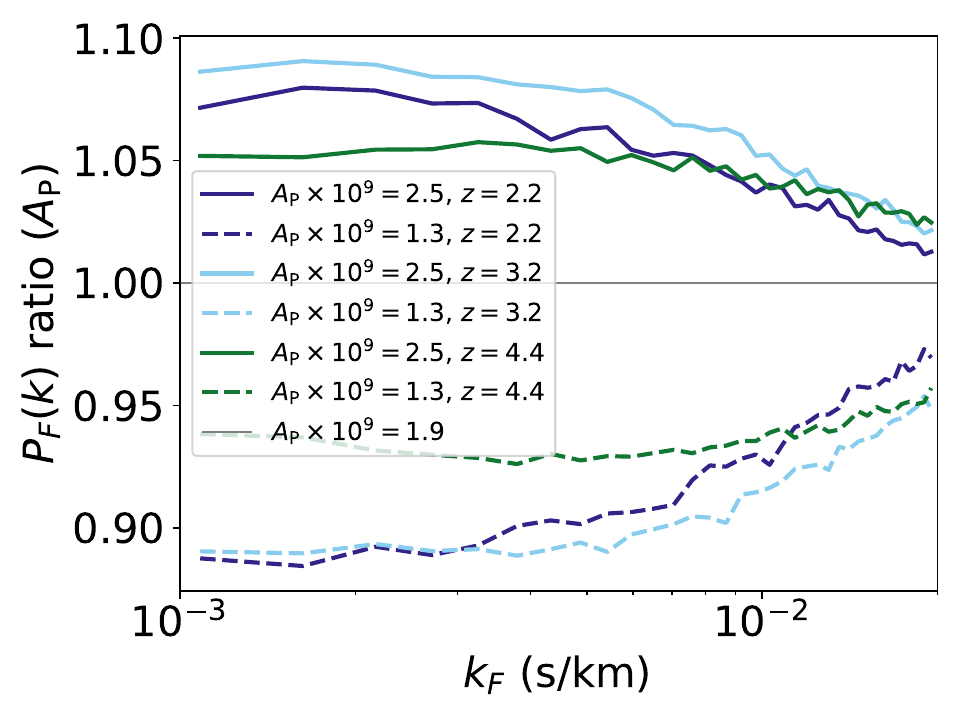}
		\includegraphics[width=0.48\columnwidth]{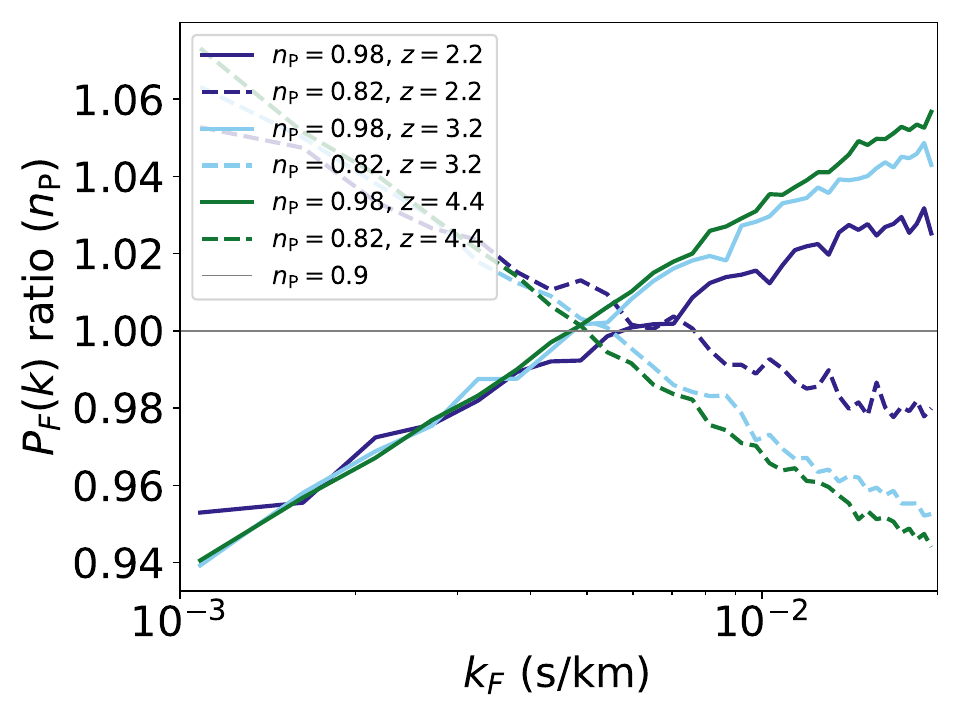} \\
    \includegraphics[width=0.48\columnwidth]{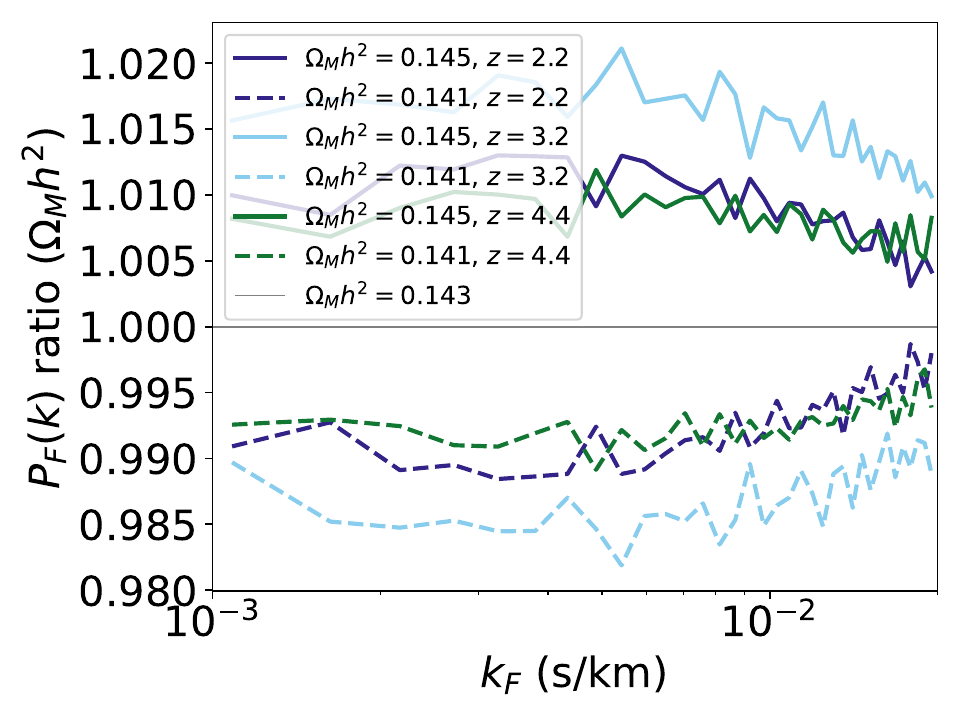}
    	\includegraphics[width=0.48\columnwidth]{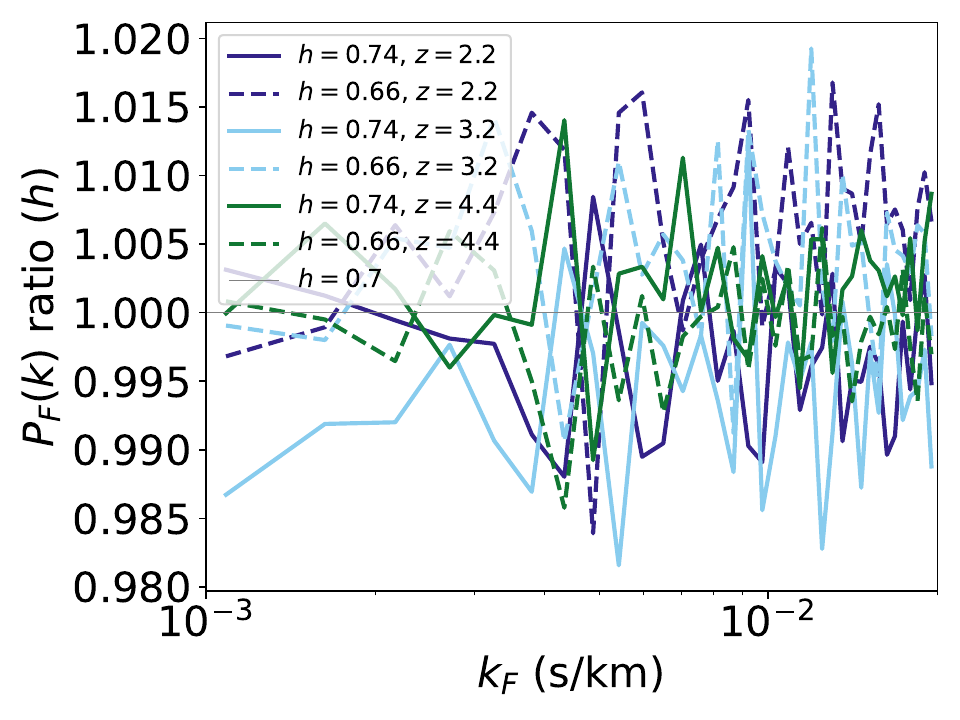}
    \caption{Predicted effect on the flux power spectra from the full multi-fidelity emulator when changing a single simulation parameter, for three representative redshifts. All other parameters are fixed to the midpoint of their range. Shown is the ratio of the 1D flux power spectrum between the central parameter value (the solid grey line in the legend) and the value shown in the legend. The top row shows the mean flux model parameters $\tau_0 = 1.0$ (Top-Left) and $d\tau_0 = 0.0$ (Top-Right). The other rows show the cosmological parameters $A_p = 1.9\times 10^{-9}$ (Mid-Left) and $n_P = 0.9$ (Mid-Right), $\Omega_m h^2 = 0.143$ (Bottom-Left) and $h = 0.7$ (Bottom-Right). }
    \label{fig:Apnsfluxpower}
\end{figure}

\begin{figure}
    \centering
    \includegraphics[width=0.48\columnwidth]{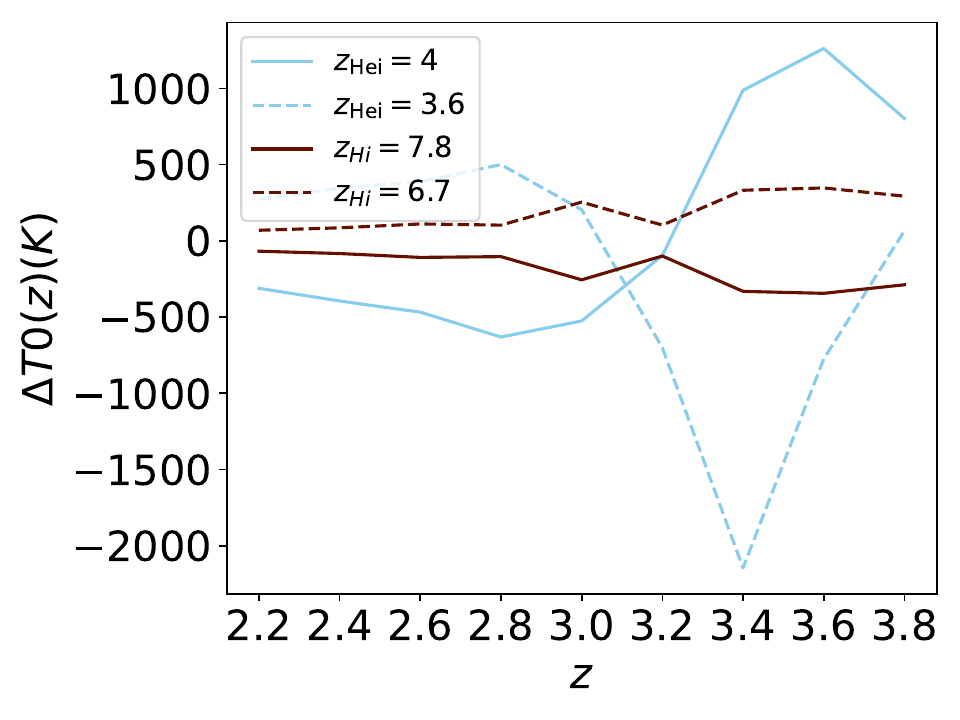}
    \includegraphics[width=0.48\columnwidth]{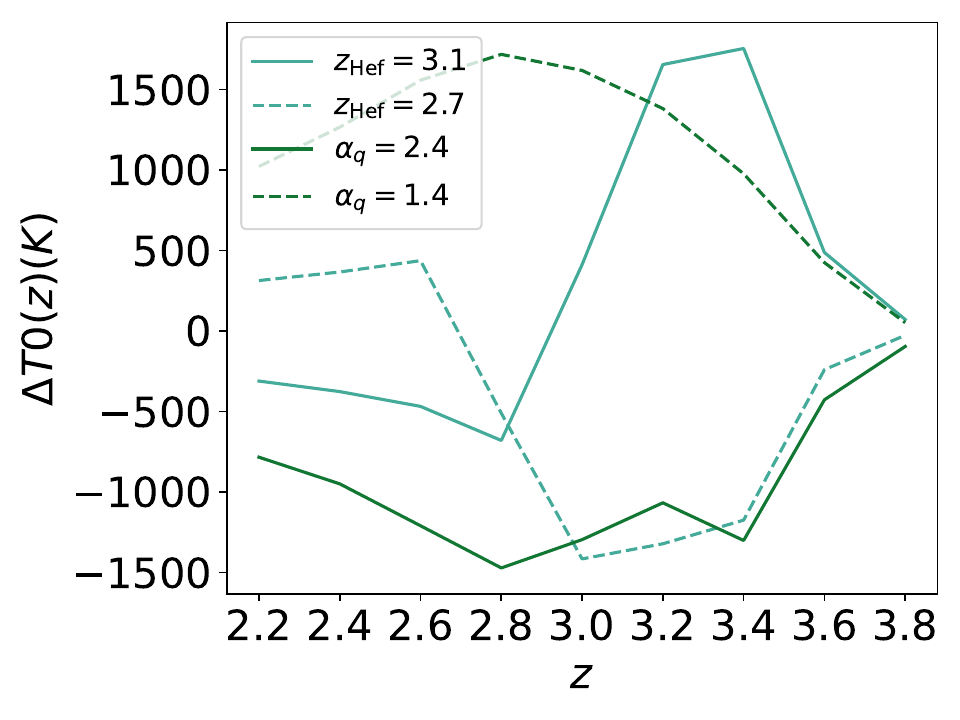} \\
    \caption{Effect on the mean IGM temperature $T_0$ as a function of redshift, for the four reionization parameters which significantly change $T_0$. Shown is the change in $T_0$ between the central parameter value and the value shown in the legend. The reference values are $z_\mathrm{Hei} = 3.8$,  $z_\mathrm{Hef} = 2.9$, $\alpha_q = 1.9$ and $z_\mathrm{HI} = 7.25$ in both panels. (Left) Changes to $z_\mathrm{Hei}$ and $z_\mathrm{HI}$. (Right) Changes to $z_\mathrm{Hef}$ and $\alpha_q$.}
    \label{fig:meantempfluxpower}
\end{figure}

\begin{figure}
    \centering
	\includegraphics[width=0.48\columnwidth]{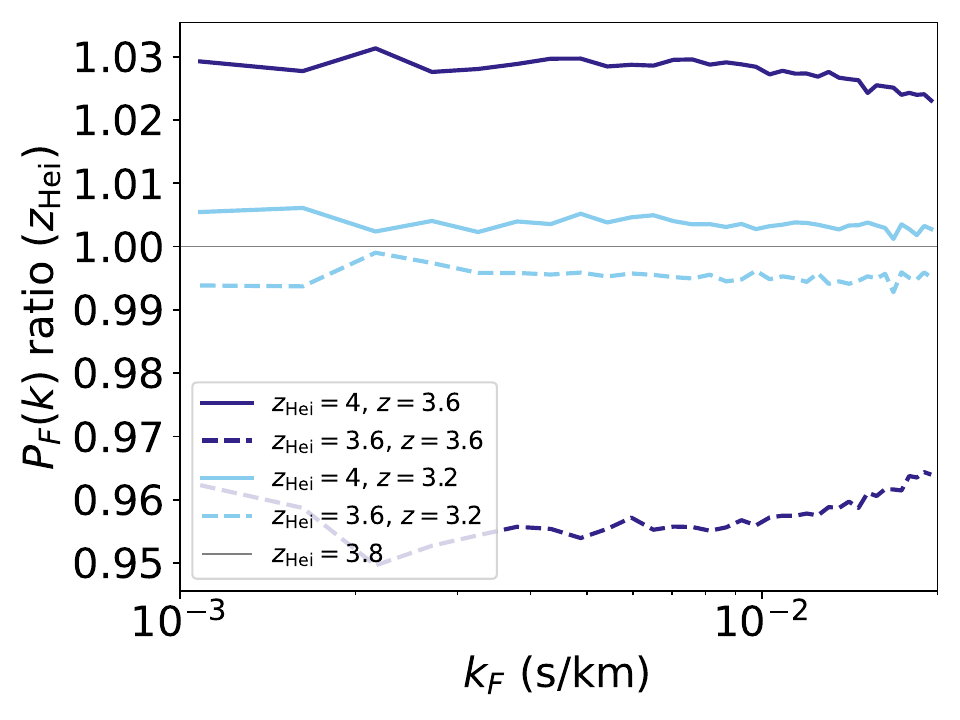}
    \includegraphics[width=0.48\columnwidth]{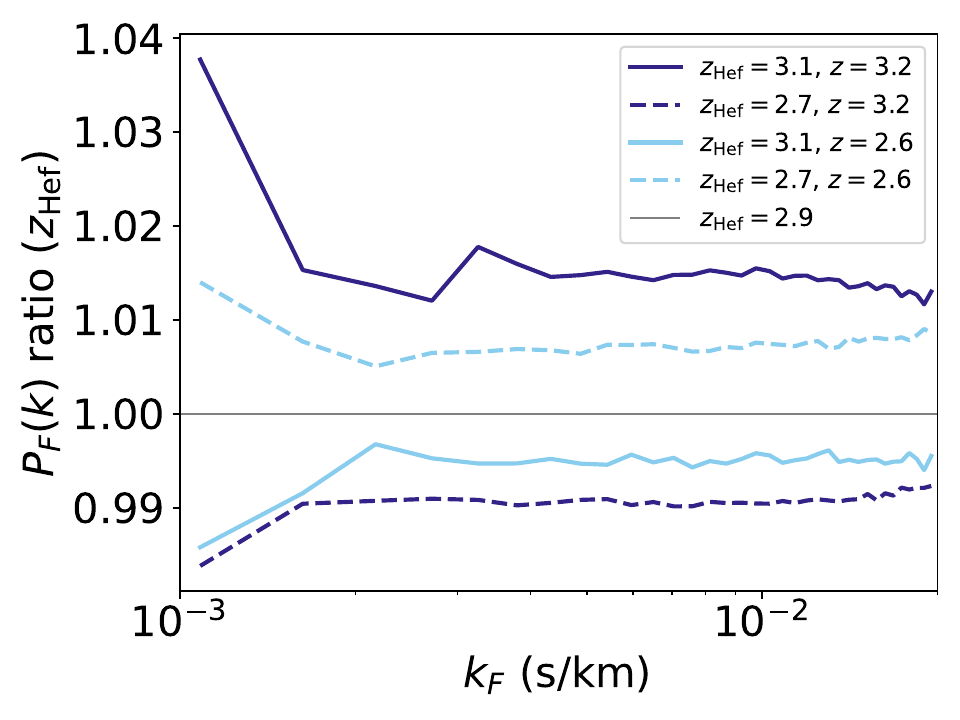} \\
	\includegraphics[width=0.48\columnwidth]{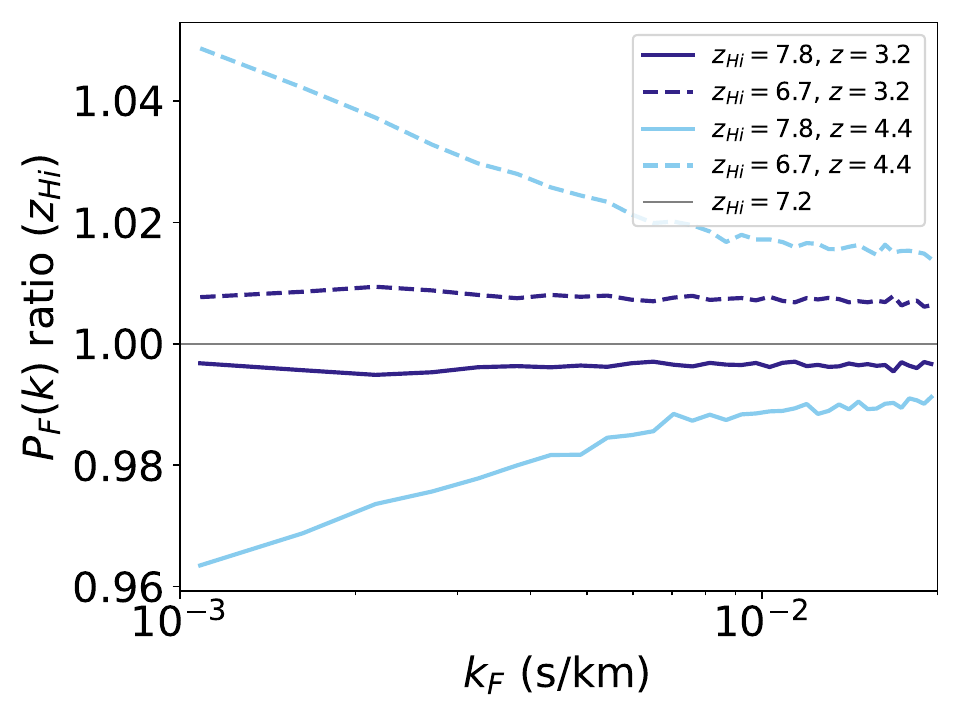}
    \includegraphics[width=0.48\columnwidth]{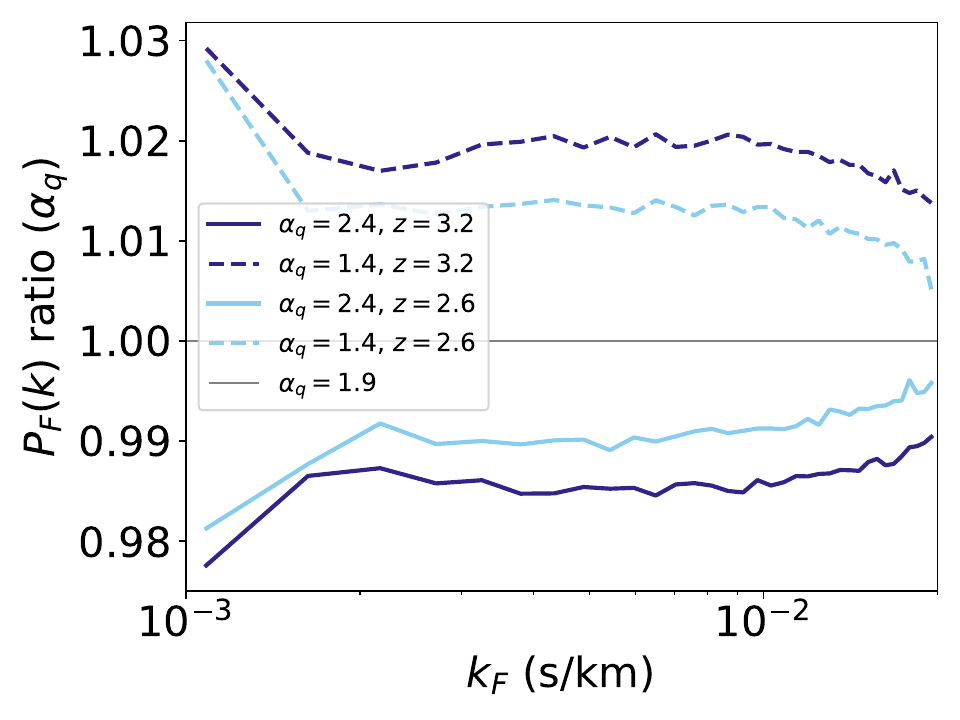}
    \caption{Predicted effect on the flux power spectra from the full multi-fidelity emulator when changing a single simulation parameter, for three representative redshifts. All other parameters are fixed to the midpoint of their range, which is given as the grey line in the legend. Shown is the ratio of the 1D flux power spectrum between the central parameter value and the value shown in the legend. Shown are the four reionization parameters $z_\mathrm{Hei} = 3.8$ (Top-Left) and $z_\mathrm{Hef} = 2.9$ (Top-Right), $z_\mathrm{HI} = 7.25$ (Bottom-Left) and $\alpha_q =1.9$ (Bottom-Right).}
    \label{fig:zhefluxpower}
\end{figure}

In this Section we discuss the effect on the flux power spectrum and mean IGM temperature predicted by our multi-fidelity emulator from varying each simulation parameter individually. We can use this to understand which parameters are most strongly measured by the \Lya~forest, and where the strongest degeneracies lie. Figure~\ref{fig:Apnsfluxpower} shows the mean flux and cosmological parameters. The top row shows the mean flux rescaling parameters $\tau_0$ and d$\tau_0$, which control the total abundance of neutral hydrogen as a function of redshift. As a reminder, these parameters are defined so that  $\tau_0 = 1$ and d$\tau_0 = 0$ correspond to $\tau_\mathrm{eff} = 0.0023 (1+z)^{3.65}$. The mean flux controls the amplitude of the flux power spectrum, and has an effect analogous to the linear bias parameter of galaxy surveys. A good order of magnitude model is that the amplitude of the flux power spectrum is proportional to $\tau_0$, with increased sensitivity at high redshift where the overall absorption is larger. Note that, by construction, the effect of d$\tau_0$ at $z=3$ is zero.

Importantly, because the mean flux parameter must be constrained first and is measured by the amplitude of the flux power spectrum, the \Lya~forest measures other cosmological parameters only insofar as they change the power spectrum shape or redshift evolution. The initial power spectrum  amplitude $A_P$, shown in the middle row of Figure~\ref{fig:Apnsfluxpower}, also changes the flux power spectrum amplitude. However, it is not degenerate with the mean flux. Increasing $A_P$ increases the flux power spectrum amplitude more on large scales, with small scales partially washed out by nonlinear growth. The spectral index $n_P$ (right panel of middle row) changes the slope of both the initial matter power spectrum and the flux power spectrum, but again is partially washed out on small scales for $z < 3$.

The above four parameters dominate the information content of the flux power spectrum on these scales \cite{Pedersen:2022}. The other two cosmological parameters are the growth function $\Omega_M h^2$ and $h$, whose effect on the flux power spectrum is significantly smaller. As expected, for constant $\Omega_M h^2$, changing $h$ also changes $\Omega_M$ and thus slightly shifts the mapping between physical and velocity units (the shift is slight because at $z=2$ we are deep in the matter dominated era). The small, poorly correlated, change in the flux power spectrum is due mostly to modes shifting between bins and thus is dominated by the small residual effect of cosmic variance. The growth function $\Omega_M h^2$ changes the amplitude of the flux power spectrum in a similar but smaller way to $A_P$, as the growth function changes the growth rate between the early universe and $z = 4.6$. $A_P$ is already fairly well constrained by Planck \cite{Planck:2018}, and so our emulator limits are relatively tight compared to the constraints available from the \Lya~forest alone.

Figure~\ref{fig:meantempfluxpower} shows the effect of the thermal parameters on the mean IGM temperature.
A later hydrogen reionization redshift gives the gas less time to cool after the temperature boost and thus increases the temperature of the IGM moderately, although the effect is washed out once helium reionization starts. The effect of the onset of helium reionization is large during the initial phases of helium reionization, but quickly diminishes by $z=3.0$. The sign of the effect is reversed at lower redshift: a later starting point of helium reionization leads to a higher final temperature as reionization proceeds faster and the gas experiences less cooling during the reionization process. The quasar spectral index $\alpha_q$ directly controls the temperature during reionization: a lower $\alpha_q$ leads to a higher peak temperature, while a higher $\alpha_q$ leads to a lower temperature. After reionization finishes, the cooling rate is the same irrespective of the peak temperature, so reducing $\alpha_q$ still boosts the temperature at $z=2.2$. During helium reionization, the effect of  $z_\mathrm{Hef}$ is similar to $\alpha_q$. A shorter helium reionization means the gas is heated more strongly while reionization is ongoing. However, once helium reionization has finished the effect is much smaller. As with $z_\mathrm{Hei}$, a shorter duration of helium reionization leads to a higher final temperature as the gas has less time to cool.

Figure~\ref{fig:zhefluxpower} shows the effect of the thermal parameters on the flux power spectrum. Changing the gas temperature changes the neutral fraction and thus the amplitude of the flux power spectrum. This is not completely degenerate with the mean optical depth parameters as the redshift dependence is proportional to the way each parameter affects the IGM temperature, shown in Figure~\ref{fig:meanigmtempdens}. Thus the amplitude shifts are largest during helium reionization and smaller by $z=2.2$. However, since $z_\mathrm{Hei}$, $z_\mathrm{Hef}$ and $\alpha_q$ all affect the flux power spectrum amplitude in similar ways, constraints on the thermal history from the flux power spectrum alone suffer from a three way degeneracy. The best constraints will continue to come from scales which resolve the thermal cutoff in the power spectrum \cite{Gaikwad:2021}.

On large scales there is a scale-dependent effect visible when changing $z_{Hef}$ and $\alpha_q$ on the largest scales probed by BOSS. A scale-dependent bias signature is expected from patchy helium reionization, as discussed in Refs.~\cite{Pontzen:2014a, Pontzen:2014b,  Gontcho:2014}. Essentially, this effect arises because different regions of the box have different temperatures, depending on when they reionized, and these temperatures are correlated over scales of the size of a helium reionization bubble, $20$ h/Mpc.

At $z=4.4$, for a low hydrogen reionization midpoint, the emulator predicts a similar large-scale effect from our patchy hydrogen reionization model. This would not yet be detectable in BOSS data as the statistical error at this redshift is still large compared to the effect. By $z=3.2$, the effect of the hydrogen reionization midpoint has mostly disappeared. A similar effect was examined in Ref.~\cite{Molaro:2022} using radiative transfer simulations, with similar results.

\subsubsection{Effect of AGN Feedback Strength}
\label{sec:agnresult}

\begin{figure*}
\centering
\includegraphics[width=0.48\columnwidth]{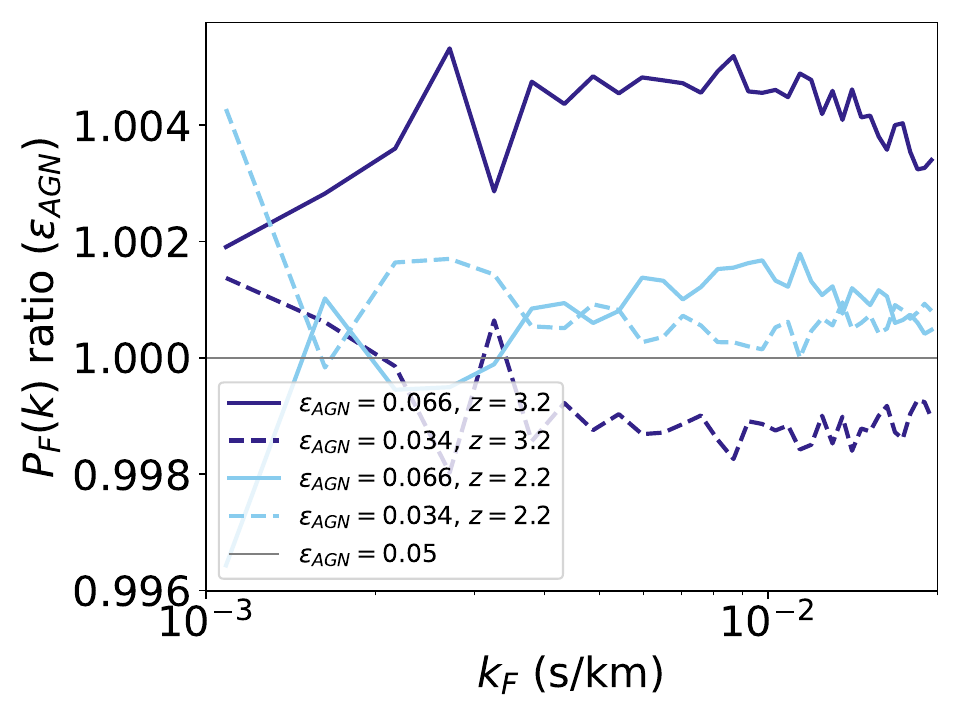}
\includegraphics[width=0.48\textwidth]{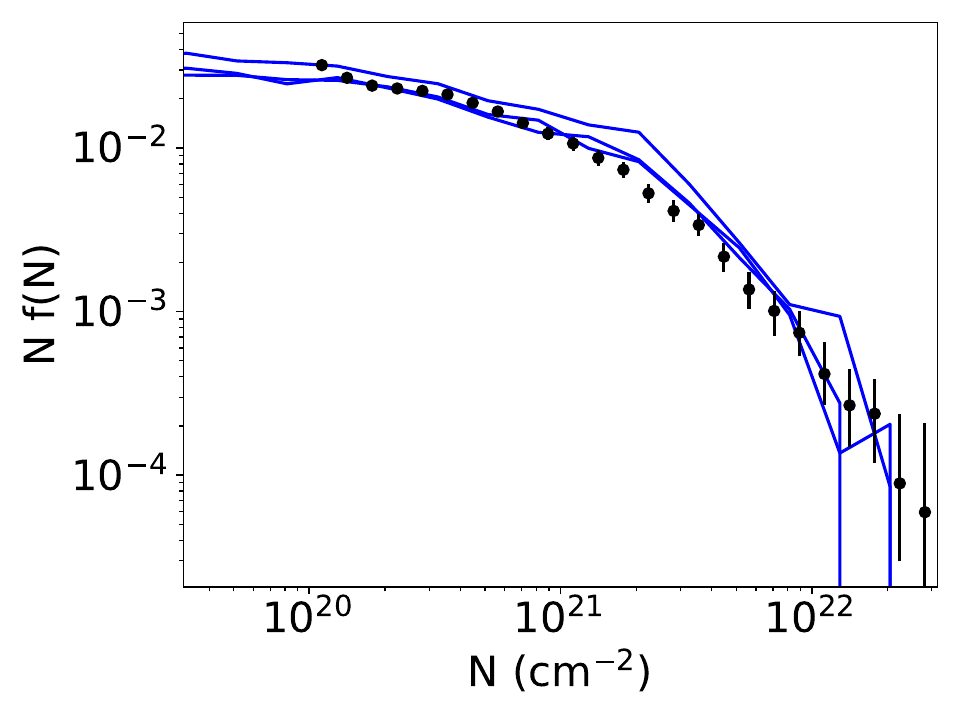}
 \caption{(Left) Predicted effect on the flux power spectra from the full multi-fidelity emulator when changing the strength of black hole feedback, $\epsilon_\mathrm{AGN}$ for $z=3.2$ and $z=2.2$. All other parameters are fixed to the midpoint of their range. Shown is the ratio of the 1D flux power spectrum between the central parameter value ($\epsilon_\mathrm{AGN} = 0.05$) and the values shown in the legend.  (Right) Column density distribution function (CDDF) of high column density absorbers from our $3$ high resolution simulations at $z=2.2$, compared to the DLA catalogue of Ref.~\cite{2021MNRAS.507..704H}, derived from SDSS DR16.}
 \label{fig:DLACDDF}
\end{figure*}

Figure~\ref{fig:DLACDDF} shows the effect of varying our AGN feedback parameter on the 1D flux power spectrum. This effect is extremely small, $<1\%$ on the BOSS scales, and shows no clear dependence on redshift or scale. Note this is not the effect of AGN feedback on the \Lya~forest, as all our simulations include AGN feedback. Instead our AGN feedback strength parameter measures the amount at which black hole accretion heats the surrounding gas. In practice this parameter most strongly controls the rate of black hole accretion: a lower value of the parameter allows the black hole to grow more before feedback shuts off the gas supply.

In retrospect it would have been sufficient to fix the AGN feedback strength to the default instead of varying it in our emulator. We did not because an initial test using a suite of $20$ Mpc/h small box simulations showed that AGN feedback strength changed the temperature of the gas. This turned out to be an artifact of sample variance in our small test box. The small box contained a small number of large mass halos hosting AGN, and so the gas temperature in the box was affected when any one of them entered a strongly accreting phase. In larger boxes individual halos are only able to affect a small fraction of the volume.


Ref.~\cite{Viel:2013} showed a maximum effect of AGN feedback on the \Lya~forest of $10\%$ at $z=2$, while Ref.~\cite{Chabanier:2020} found $8\%$ at $z=2$ and $k = 0.005$ s/km. Ref.~\cite{Chabanier:2020} uses the Horizon-AGN simulation \cite{Dubois:2016}. The AGN feedback model has two modes, one thermal and one kinetic, with the kinetic mode dominating at low accretion rates. Ref.~\cite{Viel:2013} uses the OWLS simulation suite, and delivers AGN feedback energy thermally, once black hole accretion has accumulated enough energy to heat a nearby gas particle to $10^8$ K. Despite the different implementations, the two codes agree reasonably well as to the effect of AGN feedback on the \Lya~forest.


We do not enable the ASTRID kinetic black hole feedback model \cite{Ni:2023}. Black holes in a low accretion state (as measured by a fraction of the Eddington rate) and with a mass above $M_\mathrm{BH, pivot} = 5\times 10^8 M_\odot/h$ produce a strong kinetic wind that drives gas outflows from a host galaxy cluster and so allows the simulation to match the observed cluster baryon fraction.
Our three high fidelity simulations contain $0, 1$ and $16$ black holes with masses over $5\times 10^8 M_\odot/h$ at $z=2.2$, suggesting that the potential effect of kinetic feedback on our simulations is small. For a lower mass threshold of $10^8 M_\odot/h$, as used in Illustris-TNG \cite{Weinberger:2017}, we have $5, 7, 62$ black holes. The maximum radius at which kinetic feedback could affect the gas is $\sim 1$ Mpc. Thus, even in the most generous formulation it is unlikely that kinetic feedback could affect more than $10^{-4}$ of our simulation volume. The AGN feedback model used in the SIMBA simulations \cite{SIMBA} stronger than the one in Illustris-TNG and Ref.~\cite{Tillman:2022} showed it has a substantial effect on the \Lya~forest at $z < 0.5$. However, at $z > 2$ the effect on the \Lya~forest is similar to Illustris-TNG and ASTRID \cite{Tillman:2023}.


\subsection{High Column Density Systems}
\label{sec:dlas}

Figure~\ref{fig:DLACDDF} shows the predicted column density distribution function for high column density absorbers from our three high fidelity simulations. We have generated spectra at random positions until we have $4000$ spectra, each containing a system with column density $N_{HI} > 10^{18}$ cm$^{-2}$. There is good agreement between the simulations and the observed CDDF from Ref.~\cite{2021MNRAS.507..704H}, justifying our choice to fix the supernova wind model in the simulation suite.

Even though the supernova wind model is fixed, one of the high fidelity simulations has noticeably more DLAs than the others. This may indicate the presence of some cosmological information in the DLA abundance, due to their status as tracers of low-mass halos. However, they are sensitive to the parameters of the supernova feedback \cite{Bird:2014}, which would likely foil any attempt to extract this information. This would also complicate any flux power spectrum analysis which did not mask DLAs.

\section{Conclusions}
\label{sec:conclusions}

We present the PRIYA simulation suite, a new, large, suite of cosmological simulations of the \Lya\ forest. We have used the multi-fidelity technique from \cite{Ho:2022, Fernandez:2022} to combine simulations at different resolutions, and build a cosmological emulator covering the parameter space spanned by the simulations. We include two cosmological parameters, the spectral slope $n_P$ and the perturbation amplitude $A_P$. We also vary $h$ and the growth factor $\Omega_M h^2$. These last two have a relatively small effect on the \Lya~forest on these scales, but we include them to ensure that the uncertainty resulting from them can be properly marginalised over. We include astrophysical parameters for changes in the mean flux (in post-processing), for the strength of AGN feedback, the start and end redshifts of helium reionization, the heating rate during helium reionization, and the redshift of hydrogen reionization.

The PRIYA simulations are some of the largest ever performed in a Latin Hypercube suite. We have a $120$ Mpc/h box, enough for a fair cosmological sample with minimal cosmic variance at $z=2$. Our simulations have $1536^3$ and $3072^3$ particles and thus mean interparticle spacings of $78$ and $39$ kpc/h. Our highest resolution simulations thus have 5\% higher resolution than the Illustris \cite{Vogelsberger:2014} simulation and 10\% higher resolution than the ASTRID \cite{Bird:2022, Ni:2021} simulation. We have performed $48$ low resolution simulations and $3$ high resolution simulations, which is sufficient to reduce average emulation error to the percent level, as demonstrated by our leave-one-out analysis. We required, on average, $7,000$ node-hours for each low resolution simulation and $90,000$ node-hours for each high resolution simulation. Individual simulations cost up to a factor of $2$ more or less, depending on the value of $A_P$ simulated. We computed the predicted single-parameter variations in the 1D flux power spectrum for each cosmological and astrophysical parameter, and showed that these agree with physical expectations or literature predictions.

We showed that our simulation suite can bracket the observed thermal history of the IGM, via our model for helium reionization. Note that the thermal history is measured by comparing simulations to observed statistics of high resolution quasar spectra, such as the 1D flux power spectrum for $ k_F \sim 0.01$ - $0.1$ s/km. Since our simulations provide alternative models of the flux power spectrum, in future work we may make direct constraints on our model via the small-scale flux power spectrum, rather than through the derived thermal history. We show that the effect of patchy helium reionization is imprinted into the 1D flux power spectrum on large scales. An excess appears around the scale corresponding to the size of the helium reionization bubble, at its largest near the end of reionization. We also detect a large-scale enhancement connected with the inhomogeneity of hydrogen reionization. This is distinct from the effect of helium reionization by being visible at higher redshift, and by having a shallower power law with scale. Both effects are unlikely to be robustly detectable in current data, but could potentially be detected in future in order to constrain inhomogenous reionization models.

Our simulations include a full physics galaxy formation model with stellar and AGN feedback. We fix the stellar feedback model and justify this by showing we reproduce the observed column density function of DLAs. We show that, although the presence of AGN feedback affects the
\Lya~forest, varying the strength of thermal AGN feedback has a small effect. The black holes simply accrete more gas and deposit a similar amount of energy into feedback. Our simulation suite is the first to self-consistently include AGN feedback, rather than using a correction function. This will allow us to re-use our simulation suite for other applications in future work. We include a self-shielding prescription and mask out the resulting DLAs, as is done in the observational pipeline. Thus, dense galactic gas in our simulations does not affect the flux power spectrum.

In a followup work, we have developed a likelihood function to compare the predictions from our emulator to the 1D flux power spectrum from BOSS. We will use this likelihood function to place posterior constraints on cosmological parameters using existing BOSS and future DESI data. Much of the new information in future surveys will come from statistics other than the 1D flux power spectrum. The correlation function between quasars was measured by Ref.~\cite{Slosar:2011}, and has been used to detect the baryon acoustic oscillation (BAO) feature and thus constrain the expansion rate at $z\sim 2.3$ \cite{dSAgathe:2019, Cuceu:2022}. Concurrently, \Lya~tomography surveys with a high sightline density have allowed mapping coherent Mpc-scale overdensities \cite{Lee:CLAMATO, LATIS, Qezlou:2022, Horowitz:2022}. The simulation suite is applicable to other cosmological probes, in particular the emerging field of line intensity mapping \cite{Kovetz:2017}, and its cross-correlation with the \Lya~forest \cite{Qezlou:2023}. In future work we will use the simulation suite and spectra presented here for comparisons to these other summary statistics.

\section*{Acknowledgements}
MAF was supported by a National Science Foundation Graduate Research Fellowship under grant No. DGE-1326120. MFH is supported by a National Aeronautics and Space Administration FINESST under grant No. ASTRO20-0022. SB was supported by NSF grant AST-1817256 and NASA-80NSSC21K1840. MQ is supported by NSF grant AST-2107821.

Computing resources were provided by Frontera LRAC AST21005.
The authors acknowledge the Frontera computing project at the Texas Advanced Computing Center (TACC) for providing HPC and storage resources that have contributed to the research results reported within this paper.
Frontera is made possible by National Science Foundation award OAC-1818253.

\section*{Data Availability}

Flux power spectra generated from the low and high resolution simulations, as well as the trained emulators, will be made available on publication. The simulation software is available at \url{https://github.com/MP-Gadget/MP-Gadget}. The software used to create the spectra is available at \url{https://github.com/sbird/fake_spectra}. The trained emulator, flux power spectrum and emulator software is at \url{https://github.com/sbird/lya_emulator}. Scripts to create the figures are available at \url{https://github.com/sbird/lya_suite_paper.git}. $3.4$ TB of mock spectra are available upon request.

\bibliographystyle{JHEP}
\bibliography{refs}

\appendix

\label{lastpage}
\end{document}